\documentclass[%
 aip,
 amsmath,amssymb,
 reprint,%
]{revtex4-1}

\usepackage{graphicx}
\usepackage{dcolumn}
\usepackage{bm}

\usepackage[utf8]{inputenc}
\usepackage[T1]{fontenc}
\usepackage{mathptmx}
\usepackage{etoolbox}
\usepackage{subcaption}
\usepackage{color}

\makeatletter
\def\@email#1#2{%
 \endgroup
 \patchcmd{\titleblock@produce}
  {\frontmatter@RRAPformat}
  {\frontmatter@RRAPformat{\produce@RRAP{*#1\href{mailto:#2}{#2}}}\frontmatter@RRAPformat}
  {}{}
}%

\let\vec\mathbf
\DeclareMathOperator{\Tr}{Tr}

\makeatother

\graphicspath{{./graphics/}}

\begin{document}

\preprint{AIP/123-QED}

\title[DDFT with Inertia and Background Flow]{Dynamic Density Functional Theory with Inertia and Background Flow}
\author{R. D. Mills-Williams}%
\affiliation{Edinburgh Designs Ltd, 27 Ratcliffe Terrace, Edinburgh, United Kingdom, EH9 1SX, United Kingdom}%
\author{B. D. Goddard}
\affiliation{School of Mathematics and Maxwell Institute for Mathematical Sciences, University of
Edinburgh, EH9 3FD, United Kingdom}
\author{A. J. Archer}
\affiliation{Department of Mathematical Sciences, Loughborough University, Loughborough, Leicestershire LE11 3TU, United Kingdom}

\date{\today}

\begin{abstract}
We present dynamic density functional theory (DDFT) incorporating general inhomogeneous, incompressible, time dependent background flows and inertia, describing externally driven passive colloidal systems out of equilibrium. We start by considering the underlying nonequilibrium Langevin dynamics, including the effect of the local velocity of the surrounding liquid bath, to obtain the nonlinear, nonlocal partial differential equations governing the evolution of the (coarse--grained) density and velocity fields describing the dynamics of colloids. Additionally, we show both with heuristic arguments, and by numerical solution, that our equations and solutions agree with existing DDFTs in the overdamped (high friction) limit. We provide numerical solutions that model the flow of hard spheres, in both unbounded and confined domains, and compare to previously--derived DDFTs with and without the background flow.
\end{abstract}

\maketitle

\section{Introduction}\label{sec:intro}

Statistical mechanical methods such as dynamic density functional theories (DDFTs) are increasingly popular approaches to model out of equilibrium colloidal systems. Having been applied to a multitude of systems, DDFTs have been shown to accurately describe many nonequilibrium fluid properties, such as: phase separation in binary fluids\cite{archer2005dynamical}, inertia, with\cite{goddard2012unification} and without\cite{archer2009dynamical} hydrodynamic interactions (HI), orientation\cite{WittkowskiLowen11, duran2016dynamical}, and thermal fluctuations on the hydrodynamic scale\cite{duran2017general}. Such methods predict the evolution of the one-body density distribution out of equilibrium with remarkable accuracy when compared to the underlying stochastic (Langevin) equations\cite{hansen2013theory, te2020classical}. Most existing DDFTs, however, are derived from thermostated equilibrium molecular dynamics equations, such as Langevin's equations, in a \emph{static} heat bath, and therefore do not allow for the bath in which the colloidal particles are immersed to be flowing. This limitation for existing models of nonequilibrium colloidal systems is due to a central assumption in deriving the Langevin equations: that the atoms or molecules which make up the heat bath undergo Hamiltonian ballistic dynamics, exchanging momentum with the larger colloidal particles via elastic collisions. This is correct in so far as the resulting dynamics satisfies the fluctuation--dissipation relation \cite{callen1951irreversibility} for static baths. But for a general nonequilibrium system with a given nonzero background flow, a Hamiltonian description of the bath will not produce the correct dynamics or fluctuation--dissipation relation. This has been shown rigorously by \citet{dobson2013derivation} and is due to that fact the finite local strain rate on the colloidal particles between bath atom/molecule collisions is neglected under a Hamiltonian description of the bath dynamics.

Elsewhere, classical Nonequilibrium Molecular Dynamics (NEMD) has been applied to colloidal dispersions driven via inhomogeneous boundary-driven flows, for example Newton's equations combined with Lees--Edwards boundary conditions (BCs) in the case of shear flow \cite{j2007statistical} or Kraynik--Reinelt BCs in the case of elongation flow \cite{todd1999new}. These methods however are limited to periodic problems, making more general NEMD methods increasingly desirable. Several strategies have been proposed to sample molecular systems with general, inhomogeneous background flows, including the SLLOD and g-SLLOD equations of motion for cold systems\cite{tuckerman1997modified, todd2007homogeneous} with proposals for thermostats (Nos{\' e}--Hoover, Gaussian isokinetic) for temperature control at high strain rate. However, such thermostats are problematic for, e.g., high strain rate planar-shear-flow with S-shaped streaming velocity profiles. Being kinetic thermostats, they interpret deviations away from linearity as excess thermal energy leading to an erroneous ordering effect, whereby for a simple atomic fluid, the system displays an accentuated alignment to the velocity streamlines \cite{todd2007homogeneous}. Additionally, Nos{\' e}--Hoover dynamics can be non-ergodic\cite{legoll2007non}.

In the deterministic DDFT continuum theory setting, there have been several previous approaches to study driven particle systems. Some of the earliest of these considered the overdamped dynamics of the solvent particles in a polymer solution, and their hydrodynamic interactions. \citet{penna2003dynamic} determined a DDFT accounting for the drifting effect of a single colloid particle in the case of both a non-interacting and an interacting polymer. \citet{dzubiella2003depletion}, and separately \citet{kruger2007colloid}, considered the case of two colloids, in order to calculate the effective depletion force of the polymer solute, the latter considering the deterministic Smoluchowski equation for the Brownian solute particles and solving it in bispherical coordinates. As has been previously observed, such models only modify the density of the solute and do not account for perturbations of the flow itself due to the solvent interacting with the colloid, essentially meaning the velocity field flows through the particle boundary.  To go beyond this, starting from the underlying Brownian dynamics, \citet{RauscherDominguezKrugerPenna07} derived a DDFT for advected colloidal particles externally driven by an inhomogeneous field field, based on a classical solution for incompressible low Reynolds number flow past a sphere. Extensions to this DDFT including HI with walls and between particles was again treated by \citet{Rauscher10} (including the formal details of the moment closure), and there have been many practical applications including: dynamics in confined geometries \cite{almenar2011dynamics}, advected colloids through DNA \cite{gutsche2008colloids}, phase transitions in microrheology \cite{reinhardt2014microrheology}, as well as the investigation of the hard sphere interactions in the presence of flow induced hydrodynamic lift in confined geometries \cite{yu2017microstructure}. The influence of solvent flow is also thought to be important in the dynamics of the density in active systems such as micro swimmers \cite{menzel2016dynamical}. However, care must be taken in the case of shear flow, since for this specific choice of flow, the advection term 
in the DDFT necessarily vanishes \cite{brader2011density}. This breakdown of the DDFT is traced back to the adiabatic assumption on the higher order ($n$-body) densities (assumed to be in equilibrium), and is corrected by considering exact solutions to the pair Smoluchowski equation for the correlation function 
in the presence of an external flow, for small flow rate \cite{scacchi2016driven}. Additionally, there have been many extensions to DDFT specific to shear flow, including: showing the importance of HI in computing the microscopic stress tensor \cite{aerov2015theory}, the investigation of laning instabilities \cite{scacchi2017dynamical}, and the emergence of nonequilibrium phase transitions depending on shear rate \cite{stopper2018nonequilibrium}.

All of the above DDFTs treat the dynamics of the colloidal particles as being overdamped (non-inertial), and there is a distinct shortcoming when it comes to the study of inertial effects in the presence of external flows. In this paper we present a generalisation of Refs.~\onlinecite{RauscherDominguezKrugerPenna07, Rauscher10, almenar2011dynamics} to include the effect of inertia and HI. We show that the theory reduces to previous DDFTs in the overdamped limit. Our DDFT is derived \emph{ab initio}, starting from the appropriately stated Langevin equations rigorously derived by \citet{dobson2013derivation}. We also present some numerical solutions demonstrating the applicability of the derived DDFT to physically motivated problems. 

\section{Nonequilibrium Langevin Dynamics}\label{sec:NLD}
Our starting point is a thermostated nonequilibrium dynamics for colloidal particles suspended in the flowing solvent (see, e.g., \citet[Equation (3.1), Section 3]{dobson2013derivation}). In particular, we consider the system of interacting stochastic differential equations (SDEs) on $\mathbb{R}^3$, which govern the positions $\vec{r}_i$ and momenta $\vec{p}_i$ of $i = 1,\ldots, N$ hard, spherically symmetric colloidal particles in the presence of a flowing (incompressible) background field of many more and much smaller particles given by
\begin{subequations}
\begin{align}
\frac{\mathrm{d}\vec{r}_i}{\mathrm{d}t} &= \frac{1}{m}\vec{p}_i, \label{eq:SDE_pos}\\
\frac{\mathrm{d}\vec{p}_i}{\mathrm{d}t} &= -\nabla_{\vec{r}_i}  V(\vec{r}^N,t) - \gamma(\vec{p}_i-m \vec{u}(\vec{r}_i,t)) +\sigma \, \vec{f}_{i}(t), \label{eq:SDE_mom}
\end{align}
\end{subequations}
where $m$ is the particle mass, $V$ is a potential energy (including one body and multi-body interactions), $\vec{u}$ is the inhomogeneous solvent flow (which may depend on both space and time),  $\gamma$ is the friction coefficient, $k_{\text{B}}$ is Boltzmann's constant, $T$ is temperature, and the constant $\sigma$ satisfies the fluctuation-dissipation relation $\sigma =  (m k_{\text{B}}T\gamma)^{1/2}$.  Additionally, $\vec{f}_i(t) = (\zeta^x_i(t),\zeta^y_i(t),\zeta^z_i(t))^\top$ is a Gaussian white noise term with mean $\langle \zeta_i^a(t)\rangle = 0$ and autocorrelation $\langle\zeta_i^a(t)\zeta_j^b(t) \rangle = 2\delta_{ij}\delta^{ab}\delta(t-t')$. 

The two conditions on the well-posedness of equations \eqref{eq:SDE_pos}--\eqref{eq:SDE_mom} in $d$ dimensions are that (i) $\vec{p}\cdot\nabla \vec{u}$ is uniformly Lipschitz continuous on $\mathbb{R}^d\times\mathbb{R}^d$ and (ii) $\nabla_{\vec{p}}\cdot (\vec{p}\cdot\nabla \vec{u}) = \Tr \nabla \vec{u} = \nabla\cdot \vec{u} = 0$. The first we regard as being mathematically reasonable and the second we note is true for incompressible background flows. Finally, by including a potential $V$, the stochastic equations \eqref{eq:SDE_pos}--\eqref{eq:SDE_mom} are equivalent to the g-SLLOD equations in the inertial reference frame at zero temperature $T = 0$, and \citet{dobson2013derivation} provide a derivation of the thermostated g-SLLOD equations from a heat bath model for a single large particle. 

We note that the stochastic equations \eqref{eq:SDE_pos}--\eqref{eq:SDE_mom} have subtle implications on the existence of steady states. Firstly, as discussed in \citet[Sec 2.4.3]{dobson2013derivation}, the stochastic equations \eqref{eq:SDE_pos}--\eqref{eq:SDE_mom} possess the invariant measure $f^{(1)}(\vec{r}_1,\vec{p}_1) \propto \exp(-(2 m k_B T)^{-1}|\vec{p}_1-m \vec{u}|^2)$ \emph{only} with the inclusion of an acceleration term $\vec{p}\cdot \nabla\vec{u}$ (arising from the flowing solvent's local strain rate) in the momentum equation. In this case however, the resulting dynamics can be seen as an inertial frame transformation of the classical static bath case (see, e.g., Ref.~\onlinecite{archer2009dynamical}). In contrast, an analytical expression for stationary states of \eqref{eq:SDE_pos}--\eqref{eq:SDE_mom}, are not known for general choice of $\vec{u}$ (other than the trivial choice $\vec{u}=0$), and hence we consider these Langevin dynamics in order to obtain non-trivial solutions not obtainable by frame transformation. Secondly, no assumptions on $\vec{u}$ (other than Lipschitz continuity and incompressibility) are required. In particular, as we will see, solvent flows with finite vorticity are permissible; this motivates us to believe that the present analysis is of particular interest since the final equations of motion are not restricted to, e.g., potential flows which may be trivially absorbed into an external one-body potential. In general we assume that $\vec{u}$ is not the gradient of some potential, and can have non-zero curl. Additionally, the angular momenta of the individual colloids are assumed negligible. To go beyond the overdamped assumption on the angular degrees of freedom of the colloids, one must couple to \eqref{eq:SDE_pos}--\eqref{eq:SDE_mom} a dynamical equation for the individual colloid angular acceleration, which has been presented before for orientable particles, albeit without external solvent flows; see e.g., Ref.~\onlinecite{duran2016dynamical}. 

\subsection*{Potential Energy}
We use the standard many-body expansion for the potential
\begin{align}
V(\vec{r}^N,t) = \sum_{i=1}^NV_1(\vec{r}_i,t) + \sum_{n=2}^{N}\frac{1}{n!}\sum_{i_1\neq \cdots \neq i_{n}=1}^N V_n(\vec{r}_{i_1},\cdots \vec{r}_{i_n}),
\end{align}
where $V_1$ is a one-body external potential and $V_n$ are the $n$-body interparticle potentials. We assume that these many-body interactions have no explicit time dependence.

\subsection*{Overdamped Dynamics}
It can be observed, by the usual heuristic arguments, that overdamped dynamics may be reproduced by assuming that the acceleration of a colloid in the flow is zero, $\mathrm{d}\vec{p}_i/\mathrm{d}t = 0$, such that
\begin{align}
0 =&-\nabla_{\vec{r}_i}  V(\vec{r}^N,t) - \gamma(\vec{p}_i-m \vec{u}(\vec{r}_i,t) ) +\sigma \, \vec{f}_{i}(t) ,
\end{align}
which, after substituting the corresponding value of $\vec{p}_i$ into \eqref{eq:SDE_pos}, gives the overdamped Langevin dynamics
\begin{align}
 \frac{\mathrm{d}\vec{r}_i}{\mathrm{d}t}  &=  \vec{u}(\vec{r}_i,t) - \frac{1}{m\gamma} \nabla_{\vec{r}^N}  V(\vec{r}^N,t) + \frac{\sigma}{m\gamma}\,\vec{f}_i. \label{eq:brownain_limit_SDE}
\end{align}
The dynamics \eqref{eq:brownain_limit_SDE} are recovered for DDFTs modelling colloids in a flowing solvent driven externally by a velocity potential in the overdamped regime, e.g., \citet{RauscherDominguezKrugerPenna07}, \citet{Rauscher10} but here without the assumption that $\vec{u}$ is the gradient of a potential. The overdamped limit may be performed more rigorously through an adiabatic elimination process that is based on a multiscale analysis applied to the Fokker-Planck equation associated to Eqs.~\eqref{eq:SDE_pos}--\eqref{eq:SDE_mom} (see Ref.~\onlinecite{goddard2012overdamped} for a version without a background flow but including hydrodynamic interactions).

\section{Towards A DDFT With External Flow}\label{sec:Towards A DDFT With External Flow}\label{sec:towards_ddft_with_flow}

The Fokker-Planck equation associated to the Langevin dynamics \eqref{eq:SDE_pos}--\eqref{eq:SDE_mom} is the partial differential equation (PDE)
\begin{align}
&\partial_t f^{(N)}(\vec{r}^N,\vec{p}^N,t) +\frac{1}{m}\sum_{i=1}^N\vec{p}\cdot \nabla_{\vec{r}_i}f^{(N)}(\vec{r}^N,\vec{p}^N,t)\nonumber\\
&\quad- \sum_{i=1}^N\nabla_{\vec{r}_i}V(\vec{r}^N,t) \cdot \nabla_{\vec{p}_i}f^{(N)}(\vec{r}^N,\vec{p}^N,t)\nonumber\\
&= \sum_{i=1}^N\nabla_{\vec{p}_i}\cdot\Big[ \gamma \Big(\vec{p}_i-m \vec{u}(\vec{r}_i,t) + m k_{\text{B}}T \nabla_{\vec{p}_j} \Big) f^{(N)}(\vec{r}^N,\vec{p}^N,t)\Big] \label{eq:fokker-planck} 
\end{align}
where $f^{(N)}(\vec{r}^N,\vec{p}^N,t)$ is the probability density of finding each particle $i$ at position $\vec{r}_i$ with momentum $\vec{p}_i$ at time $t$, referred to as the $N$-body density. Equation \eqref{eq:fokker-planck} is a high dimensional PDE; indeed the number of discretisation points in a standard numerical scheme, such as finite difference or pseudospectral methods, increases exponentially in $N$. Hence it is numerically intractable for all but a small number of particles.

To continue, we integrate \eqref{eq:fokker-planck} and derive equations for its moments, giving an infinite hierarchy to be truncated with moment closure. This passage is well discussed; see e.g.\ Refs.~\onlinecite{archer2009dynamical, goddard2012unification} for an in depth derivation of the moment closure. Here we provide only the essential details in order to connect the main quantities and equations; see Ref.~\onlinecite{goddard2012unification} for relevant definitions. Multiplying \eqref{eq:fokker-planck} by $N$ and integrating over all $\vec{p}_i$ and all but one particle position, $\vec{r}_1$, yields the continuity equation \eqref{eq:cont_eqn} for the one body density $\varrho\equiv \varrho^{(1)}$ given by
\begin{align}\label{eq:cont_eqn}
\partial_t\varrho(\vec{r}_1,t)+\nabla_{\vec{r}_1}\cdot \vec{j}(\vec{r}_1,t)= 0,
\end{align}
where  $\varrho(\vec{r}_1,t)$ is the one body density (Eq.~\eqref{eq:one_body_density_int_eqn} below, with $n=1$) and the current $\vec{j}$ is defined by
\begin{align}
\vec{j}(\vec{r}_1,t):= \int\mathrm{d}\vec{p}_1\,\frac{\vec{p}_1}{m}f^{(1)}(\vec{r}_1,\vec{p}_1,t).
\end{align}

Multiplying \eqref{eq:fokker-planck} by $N\vec{p}_1/m$ and once again integrating over all variables except for $\vec{r}_1$ yields the current evolution equation
\begin{align}
&\partial_t\vec{j}(\vec{r}_1,t)+\nabla_{\vec{r}_1}\cdot \int \mathrm{d}\vec{p}_1\, \Big(\frac{\vec{p}_1}{m}\otimes \frac{\vec{p}_1}{m}\Big) f^{(1)}(\vec{r}_1,\vec{p}_1,t)\nonumber\\
&\quad +\frac{1}{m}\varrho(\vec{r}_1,t)\nabla_{\vec{r}_1}V_1(\vec{r}_1,t) \nonumber\\
&\quad+\frac{1}{m}\sum_{n=2}^N\int \mathrm{d}\vec{r}^{n-1}\nabla_{\vec{r}_1}V_n(\vec{r}^n)\varrho^{(n)}(\vec{r}^n,t)\nonumber\\
&\quad+\gamma(\vec{j}\big(\vec{r}_1,t)-\varrho(\vec{r}_1,t)\vec{u}(\vec{r}_1,t)\big) = 0,\label{eq:j_current_evolve}
\end{align}
where $\mathrm{d}\vec{r}^{n-1}$ denotes $\mathrm{d}\vec{r}^2 \ldots \mathrm{d}\vec{r}^{n}$.
To close the system \eqref{eq:cont_eqn}--\eqref{eq:j_current_evolve}, equations are needed for the quantities involving $\varrho^{(n)}(\vec{r}^n,t)$ and $f^{(1)}(\vec{r}_1,\vec{p}_1,t)$.   For the former we employ the standard adiabatic approximation\cite{marconi1999dynamic, archer2009dynamical, goddard2012unification, hansen2013theory, te2020classical, schmidt2022power}, while for the latter we use a modified local equilibrium (more properly, local steady state) approximation, which takes into account the background flow.

\subsection{Adiabatic Approximation}
We consider the free energy of the non-equilibrium system at constant temperature and introduce the Helmholtz free energy functional\cite{hansen2013theory, te2020classical, evans1979nature}
\begin{align}
\mathcal{F}[\varrho]& =k_{B}T \int \mathrm{d}\vec{r}_1\, \varrho(\log (\Lambda^3\varrho)-1)+\int\mathrm{d}\vec{r}_1\,\varrho V_1(\vec{r}_1,t)\nonumber\\
&\quad+\mathcal{F}_{ex}[\varrho].\label{eq:Helmholtz_free_energy_augmented}
\end{align}
In the above functional $\Lambda$ is the de Broglie wavelength (which is superfluous) and $\mathcal{F}_{ex}[\varrho]$ is the excess over ideal gas term which is, in general, unknown; we will later make use of an equilibrium identity (sum rule) connecting it to the many-body parts of the potential. First we introduce the reduced phase space densities, defined by
\begin{align}
f^{(n)}(\vec{r}^n,\vec{p}^n,t) &:= \frac{N!}{(N-n)!}\int \mathrm{d}\vec{r}^{N-n}\,\mathrm{d}\vec{p}^{N-n}\, f^{(N)}(\vec{r}^N,\vec{p}^N,t),\label{eq:reduced_phase_space_densities}
\end{align}
and the reduced configuration densities $\varrho^{(n)}(\vec{r}^n,t)$, given by 
\begin{align}\label{eq:one_body_density_int_eqn}
\varrho^{(n)}(\vec{r}^n,t) &:= \int\mathrm{d}\vec{p}^{n}\, f^{(n)}(\vec{r}^n,\vec{p}^n,t),
\end{align}
where, again, $\varrho^{(1)}\equiv \varrho$ and $\mathrm{d}\vec{x}^{N-n} = \mathrm{d}\vec{x}^{n+1} \ldots \mathrm{d}\vec{x}^{N}$. It is assumed that the $\varrho^{(n)}(\vec{r}^n,t)$ are well approximated by their counterpart $n$--body densities for an equilibrium fluid and we will now determine the required steady sum rule. To establish a well defined equilibrium density distribution we must assume $\vec{u} \equiv 0$, which, e.g., can be viewed as a valid initial condition for the Fokker--Planck equation \eqref{eq:fokker-planck}. In particular, at equilibrium, the one body phase space function is assumed to be a Maxwell--Boltzmann distribution 
\begin{align}\label{eq:maxwell_boltzmann_dist}
f^{(1)}(\vec{r}_1,\vec{p}_1):=\tfrac{\varrho(\vec{r}_1)}{(2\pi m k_{B}T)^{3/2}}\exp\left(-\tfrac{|\vec{p}_1|^2}{2mk_{B}T}\right). 
\end{align}
Under this assumption, we see that, as the momentum distribution equilibrates, the kinetic pressure term may be calculated by 
\begin{align}
&\nabla_{\vec{r}_1}\cdot \int \mathrm{d}\vec{p}_1\, \Big(\frac{\vec{p}_1}{m}\otimes \frac{\vec{p}_1}{m} \Big) f^{(1)}(\vec{r}_1,\vec{p}_1) = \tfrac{k_BT}{m}\nabla_{\vec{r}_1}\varrho(\vec{r}_1,t). \label{eq:kinetic_pressure_equilibrium_int}
\end{align}
Hence, at steady flow, using equations \eqref{eq:maxwell_boltzmann_dist} and \eqref{eq:kinetic_pressure_equilibrium_int}, equation  \eqref{eq:j_current_evolve} becomes
\begin{align}
& \tfrac{k_BT}{m}\nabla_{\vec{r}_1}\varrho(\vec{r}_1,t)+\frac{1}{m}\varrho(\vec{r}_1,t)\nabla_{\vec{r}_1}V_1(\vec{r}_1,t)\nonumber\\
& \quad +\frac{1}{m}\sum_{n=2}^N\int \mathrm{d}\vec{r}^{n-1}\nabla_{\vec{r}_1}V_n(\vec{r}^n)\varrho^{(n)}(\vec{r}^n,t) =0.\label{eq:j_current_evolve_equilibrium}
\end{align}
Now, by computing the gradient of the variational derivative of $\mathcal{F}$, the Euler--Lagrange equation for the steady state density is
\begin{align}
0& = \frac{1}{m}\varrho\nabla_{\vec{r}_1}\frac{\delta\mathcal{F}[\varrho]}{\delta\varrho} \nonumber\\
&= \frac{k_{B}T}{m}\nabla_{\vec{r}_1}\varrho+\frac{1}{m}\varrho\nabla_{\vec{r}_1}V_1+\frac{1}{m}\varrho\nabla_{\vec{r}_1}\frac{\delta\mathcal{F}_{\text{ex}}[\varrho]}{\delta\varrho}.\label{eq:Euler-Lagrange_eqn}
\end{align}
By subtracting \eqref{eq:j_current_evolve_equilibrium} from \eqref{eq:Euler-Lagrange_eqn} one obtains the steady sum rule\cite{evans1979nature, ArcherEvans04}
\begin{align}
\varrho(\vec{r}_1,t)\nabla_{\vec{r}_1}\tfrac{\delta\mathcal{F}_{\text{ex}}}{\delta\varrho}[\varrho]&=\sum_{n=2}^N\int\mathrm{d}\vec{r}^{n-1}\nabla_{\vec{r}_1}V_n(\vec{r}^n)\varrho^n(\vec{r}^n,t).\label{eq:excess_free_energy_equilibrium_sum}
\end{align}
We assume that \eqref{eq:excess_free_energy_equilibrium_sum} holds out of equilibrium to complete the adiabatic approximation on the higher $n$-body densities.

\subsection{Moment Closure and Local Equilibrium Approximation}
Using the definition of $\mathcal{F}$ along with the (out of equilibrium) approximation \eqref{eq:excess_free_energy_equilibrium_sum}, equation \eqref{eq:j_current_evolve} becomes
\begin{align}
&\partial_t\vec{j}(\vec{r}_1,t)+\bm{A}(\vec{r}_1,t)+\frac{1}{m}\varrho(\vec{r}_1,t)\nabla_{\vec{r}_1}\frac{\delta\mathcal{F}[\varrho]}{\delta \varrho}\nonumber\\
&\quad +\gamma\big(\vec{j}(\vec{r}_1,t)  -\varrho(\vec{r}_1,t) \vec{u}(\vec{r}_1,t) \big)=0,\label{eq:j_current_evolve_A}
\end{align}
where
\begin{align}
\bm{A}(\vec{r}_1,t) &= \nabla_{\vec{r}_1}\cdot \int \mathrm{d}\vec{p}_1\, \Big(\frac{\vec{p}_1}{m}\otimes \frac{\vec{p}_1}{m}-\frac{k_BT}{m}\bm{1}\Big) f^{(1)}(\vec{r}_1,\vec{p}_1,t).
\end{align}
Note that at steady state, $\bm{A}$ converges to zero. Away from equilibrium, we assume that the momentum of each colloid particle is distributed according to the Maxwell--Boltzmann distribution centred at the mean local velocity $\vec{v}(\vec{r}_1,t)$. We write $f^{(1)}$ as the sum of local steady and non-steady parts
\begin{align}\label{eq:ls+ns}
f^{(1)}(\vec{r}_1,\vec{p}_1,t) = f^{(1)}_{\text{ls}}(\vec{r}_1,\vec{p}_1,t)+f^{(1)}_{\text{ns}}(\vec{r}_1,\vec{p}_1,t),
\end{align}
where
\begin{align}
f^{(1)}_{\text{ls}}(\vec{r}_1,\vec{p}_1,t) = \tfrac{\varrho(\vec{r}_1,t)}{(2\pi m k_{B}T)^{3/2}}\exp\left(-\tfrac{|\vec{p}_1-m\vec{v}(\vec{r}_1,t)|^2}{2mk_{B}T}\right). 
\end{align}
The first few moments of $f^{(1)}_{\text{ls}}$ are
\begin{align}
&\int\mathrm{d}\vec{p}_1\, f^{(1)}_{\text{ls}}(\vec{r}_1,\vec{p}_1,t) = \varrho(\vec{r}_1,t),\\
&\int\mathrm{d}\vec{p}_1\,\vec{p}_1 f^{(1)}_{\text{ls}}(\vec{r}_1,\vec{p}_1,t) = m\varrho(\vec{r}_1,t)\vec{v}(\vec{r}_1,t),\label{eq:f_le_moments}\\
&\int\mathrm{d}\vec{p}_1\, |\vec{p}_1-m\vec{v}(\vec{r}_1,t)|^2f^{(1)}_{\text{ls}}(\vec{r}_1,\vec{p}_1,t) = mk_{B}T\varrho(\vec{r}_1,t)
\end{align}
and we impose the integral restrictions
\begin{align}
&\int\mathrm{d}\vec{p}_1\, f^{(1)}_{\text{ns}}(\vec{r}_1,\vec{p}_1,t) = 0,\\
&\int\mathrm{d}\vec{p}_1\,\vec{p}_1 f^{(1)}_{\text{ns}}(\vec{r}_1,\vec{p}_1,t) = 0,\label{eq:f_neq_moments}\\
&\int\mathrm{d}\vec{p}_1\, |\vec{p}_1-m\vec{v}(\vec{r}_1,t)|^2f^{(1)}_{\text{ns}}(\vec{r}_1,\vec{p}_1,t) =0.
\end{align}
By using the integral conditions \eqref{eq:f_le_moments} and \eqref{eq:f_neq_moments}, combined with the expansion \eqref{eq:ls+ns}, the continuity equation \eqref{eq:cont_eqn} becomes
\begin{align}\label{eq:cont_eqn_v}
\partial_t\varrho(\vec{r}_1,t) + \nabla_{\vec{r}_1}\cdot \left(\varrho(\vec{r}_1,t)\vec{v}(\vec{r}_1,t)\right) = 0.
\end{align}
We may also compute $\vec{A}(\vec{r}_1,t)$ explicitly as
\begin{align}
\bm{A}(\vec{r}_1,t) &=\nabla_{\vec{r}_1}\cdot\left(\varrho(\vec{r}_1,t)\vec{v}(\vec{r}_1,t)\otimes\vec{v}(\vec{r}_1,t)) \right)\nonumber\\
&\quad  +\nabla_{\vec{r}_1}\cdot \int \mathrm{d}\vec{p}_1\, \frac{\vec{p}_1\otimes \vec{p}_1}{m^2}f^{(1)}_{\text{ns}}(\vec{r}_1,\vec{p}_1,t).\label{eq:A_computed_explicitly}
\end{align}
We insert this expression for $\vec{A}(\vec{r}_1,t)$ into \eqref{eq:j_current_evolve_A} to obtain
\begin{align}
&\partial_t\left(\varrho(\vec{r}_1,t)\vec{v}(\vec{r}_1,t)\right)+\nabla_{\vec{r}_1}\cdot\left(\varrho(\vec{r}_1,t)\vec{v}(\vec{r}_1,t)\otimes\vec{v}(\vec{r}_1,t) \right)\nonumber\\
&\, + \nabla_{\vec{r}_1}\cdot \int \mathrm{d}\vec{p}_1\, \frac{\vec{p}_1\otimes \vec{p}_1}{m^2}f^{(1)}_{\text{ns}}(\vec{r}_1,\vec{p}_1,t)\nonumber\\
&\, +\frac{1}{m}\varrho(\vec{r}_1,t)\nabla_{\vec{r}_1}\tfrac{\delta\mathcal{F}[\varrho]}{\delta \varrho} + \gamma\varrho(\vec{r}_1,t)(\vec{v}(\vec{r}_1,t)-\vec{u}(\vec{r}_1,t)) =0.\label{eq:mom_evolve_before_two_body_friction}
\end{align}

Now consider the following identity, for an arbitrary, suitably well behaved vector field $\vec{z}(\vec{r}_1,t)$
\begin{align}
&\partial_t\left(\varrho \vec{z}\right) +  \nabla_{\vec{r}_1}\cdot \left(\varrho \vec{z}\otimes \vec{z}\right)\nonumber\\
&=\vec{z}\partial_t\varrho  + \varrho \partial_t \vec{z} +\varrho\left(\vec{z}(\nabla\cdot \vec{z})+ (\vec{z}\cdot \nabla)\vec{z}\right) + \vec{z}\left(\vec{z}\cdot \nabla \varrho\right). \label{eq:z_identity}
\end{align}
This result follows from the product rule of differentiation and, e.g., equation (27) of Ref.~\onlinecite{archer2009dynamical}.
By setting $\vec{z} = \vec{v}(\vec{r}_1,t)$, using the continuity equation \eqref{eq:cont_eqn_v}, and neglecting the $f_{\text{ns}}$ term, (since it can be shown to be small, at least in the overdamped limit \cite{goddard2012overdamped} $\gamma\to\infty$), we obtain our equations of motion. 
\begin{align}
\begin{cases}
&  \qquad\quad \partial_t\varrho(\vec{r}_1,t)+\nabla_{\vec{r}_1}\cdot\left(\varrho(\vec{r}_1,t)\vec{v}(\vec{r}_1,t)\right) =0,\\[2mm]
& \partial_t\vec{v}(\vec{r}_1,t)+(\vec{v}(\vec{r}_1,t)\cdot\nabla_{\vec{r}_1})\vec{v}(\vec{r}_1,t) \\
& \qquad\quad +\tfrac{1}{m}\nabla_{\vec{r}_1}\frac{\delta \mathcal{F}[\varrho]}{\delta \varrho}(\vec{r}_1,t) + \gamma\vec{w}(\vec{r}_1,t) = 0,
\end{cases} \label{eq:DDFT}
\end{align}
where $\vec{w}(\vec{r}_1,t):= \vec{v}(\vec{r}_1,t)- \vec{u}(\vec{r}_1,t)$. The momentum equation in \eqref{eq:DDFT} should be compared with the inertial DDFT given in Eq.~(30) of Ref.~\onlinecite{archer2009dynamical}. We note that the correct inertial dynamics \eqref{eq:DDFT} cannot be obtained from the static bath case therein simply by a transformation between the fixed and moving frame of reference (moving with the streaming velocity $\vec{u}$). This comes as a consequence of the nonlinear advection term in \eqref{eq:DDFT}. One can see this by transforming the static bath case in Ref.~\onlinecite{archer2009dynamical} to the moving reference frame, which is equivalent to shifting the transport of the velocity distribution $\vec{v}(\vec{r}_1,t)$ by the external flow $\vec{u}(\vec{r}_1,t)$. We see that the velocity advection transforms to
\begin{align}
D_t \vec{v}(\vec{r}_1,t) \to \partial_t\vec{w}(\vec{r}_1,t)+(\vec{w}(\vec{r}_1,t)\cdot\nabla_{\vec{r}_1})\vec{w}(\vec{r}_1,t),
\end{align}
which differs from the advection term in \eqref{eq:DDFT}. We note that it is possible, via a careful choice of $\vec{u}(\vec{r}_1,t)$, to make the steady state velocity and density distribution resulting from \eqref{eq:DDFT} and a frame--transformed Eq.~(30) in Ref.~\onlinecite{archer2009dynamical} to be made the same. Nevertheless, the passage to equilibrium will clearly not be the same, due to the difference in the nonlinear advection terms.

This concludes an \emph{ab initio} derivation of the inertial DDFT in the presence of an inhomogeneous, incompressible background flow. We reemphasise that \eqref{eq:DDFT} is not restricted to curl-free $\vec{u}$, and may therefore not be trivially derived by appropriately perturbing the one-body field $V_1$ to include a velocity potential accounting for the flowing solvent.

\subsection{Overdamped Limit }

We now consider the overdamped limit by attempting to close \eqref{eq:DDFT}, into a single equation for the dynamics of the density alone. Returning to Eqs.~\eqref{eq:cont_eqn} and \eqref{eq:j_current_evolve_A}, we have under the local-steady approximation, the conservation of momentum
\begin{align}
&\partial_t\vec{j} +\bm{A}_{ls}+\frac{1}{m}\varrho\nabla_{\vec{r}_1}\frac{\delta\mathcal{F}[\varrho]}{\delta \varrho}+\gamma(\vec{j}-\varrho\vec{u})=0, \label{eq:mom_eqn_ls}
\end{align}
where $\bm{A}_{ls}$ is given by \eqref{eq:A_computed_explicitly} with $f_{ns} = 0$. Differentiating \eqref{eq:cont_eqn} we obtain
\begin{align}
\partial^2_t\varrho + \partial_t \left(\nabla_{\vec{r}_1}\cdot \vec{j} \right) = 0.\label{eq:cont_eqn_tt}
\end{align}
Additionally, by taking the divergence of \eqref{eq:mom_eqn_ls}, we have
\begin{align}
&\partial_t \left(\nabla_{\vec{r}_1}\cdot \vec{j} \right) + \nabla_{\vec{r}_1}\cdot\bm{A}_{ls}+\tfrac{1}{m}\nabla_{\vec{r}_1}\cdot\left(\varrho(\vec{r}_1,t)\nabla_{\vec{r}_1}\tfrac{\delta\mathcal{F}[\varrho]}{\delta \varrho}\right)\nonumber\\
&\quad+\gamma \nabla_{\vec{r}_1}\cdot\vec{j} =0. \label{eq:div_mom_eqn_ls}
\end{align}
Combining Eqs.~\eqref{eq:cont_eqn_tt} and \eqref{eq:div_mom_eqn_ls}, we obtain
\begin{align}
&\partial^2_t\varrho + \gamma\left(\partial_t\varrho + \nabla_{\vec{r}_1}\cdot (\varrho \,\vec{u}) \right) -\tfrac{1}{m}\nabla_{\vec{r}_1}\cdot\left(\varrho(\vec{r}_1,t)\nabla_{\vec{r}_1}\tfrac{\delta\mathcal{F}[\varrho]}{\delta \varrho}\right)
\nonumber\\
&\quad= \nabla_{\vec{r}_1}\cdot\bm{A}_{ls}.\label{eq:combined_ls_mass_mom_eqn}
\end{align}
It remains to treat the term on the right hand side. We note that in Ref.~\onlinecite{goddard2012unification} we have $\bm{A}_{ls} = \bm{A}_{le}$, and, by the results of Ref.~\onlinecite{goddard2012overdamped}, at least for a two-body potential truncation of $\mathcal{F}_{ex}[\varrho]$, terms containing $\bm{A}_{le}$ are negligible in the high-friction limit, $\gamma \to \infty$. Hence, under a similar adiabatic elimination process, based on a multiscale analysis of the Fokker-Planck equation \eqref{eq:fokker-planck}, we expect that $\bm{A}_{ls} $ will also be negligible in the presence of the solvent flow $\vec{u}$. Therefore, \eqref{eq:combined_ls_mass_mom_eqn} maybe approximated as
\begin{align}
& \partial_t^2\varrho + \gamma\left(\partial_t\varrho + \nabla_{\vec{r}_1}\cdot (\varrho \,\vec{u}) \right) -\tfrac{1}{m}\nabla_{\vec{r}_1}\cdot\left(\varrho(\vec{r}_1,t)\nabla_{\vec{r}_1}\tfrac{\delta\mathcal{F}[\varrho]}{\delta \varrho}\right)\approx  0.\label{eq:high_friction_limit}
\end{align}
The second-order time derivative $\partial_t^2 \varrho $ can be heuristically neglected in the high-friction limit, though its effect could be important on short timescales. Hence, previous overdamped DDFTs with external flow, e.g., \citet{RauscherDominguezKrugerPenna07,kruger2007colloid,Rauscher10,almenar2011dynamics}, are recovered. Assuming $\varrho$ is strictly positive, we may also establish that the free energy is a monotonically  decreasing function of time along the solvent flow
\begin{eqnarray}
D_t \mathcal{F} &=& \int \left(\partial_t \varrho + \vec{u}\cdot \nabla\varrho\right)\tfrac{\delta \mathcal{F}}{\delta \varrho}\mathrm{d}\vec{r} \nonumber \\
&=& -\tfrac{1}{m\gamma}\int \varrho |\nabla \tfrac{\delta \mathcal{F}}{\delta \varrho}|^2\mathrm{d}\vec{r} \leq 0,
\end{eqnarray}
where we have assumed that the boundary terms vanish; this is the case, e.g., when $\rho=0$ on the boundary, or when the system is equipped with for no-flux or periodic boundary conditions. Thus, this is a generalisation of the standard monotonicity of the free energy over time in overdamped DDFT.


\section{Including Hydrodynamic Interactions}
Hydrodynamic interactions (HIs) between the colloids may be introduced by modifying the friction term in Eq.~\eqref{eq:DDFT}~\cite{rex2008dynamical, rex2009dynamical, goddard2012unification}. By introducing the friction tensor $\bm{\Gamma}\in \mathbb{R}^{3N\times3N}$, the non-equilibrium Langevin dynamics \eqref{eq:SDE_pos}--\eqref{eq:SDE_mom} becomes

\begin{subequations}
\begin{align}
\frac{\mathrm{d}\vec{r}_i}{\mathrm{d}t} &= \frac{1}{m}\vec{p}_i \label{eq:SDE_pos_HI}\\
\frac{\mathrm{d}\vec{p}_i}{\mathrm{d}t} &= -\nabla_{\vec{r}_i}  V(\vec{r}^N,t) -\sum_{j=1}^{N}\bm{\Gamma}_{ij}(\vec{r}^N)
\big(\vec{p}_j-m\vec{u}(\vec{r}_j,t) \big)\\
&\quad+ \sum_{j=1}^N\bm{B}_{ij}(\vec{r}^N)\vec{f}_{j}(t) \label{eq:SDE_mom_HI}
\end{align}
\end{subequations}
where $\bm{B} = \left( mk_{\text{B}}T\bm{\Gamma}\right)^{1/2}$ and $\bm{\Gamma}$ comprises $N^2$ positive definite resistance matrices $\bm{\Gamma}_{ij}$ given by
$
\bm{\Gamma}_{ij} = \gamma\bm{1} +\gamma\tilde{\bm{\Gamma}}_{ij}
$,
where $\bm{1}$ is the $3\times 3$ identity matrix, and $\tilde{\bm{\Gamma}}_{ij}\in\mathbb{R}^{3\times3}$. Physically we typically regard the forces introduced by $\gamma\tilde{\bm{\Gamma}}_{ij}$ as short range HIs mediated by the bath, (also known as lubrication forces) and $\gamma$ corresponds to Stokes drag at infinite particle separation. We remark that \eqref{eq:SDE_pos_HI}--\eqref{eq:SDE_mom_HI} are the natural generalisation to include HIs but we stress that these equations have not been rigorously derived. However, as we will see, they do lead to the expected set of continuum equations, consistent with \eqref{eq:DDFT} and \eqref{eq:high_friction_limit} upon taking appropriate limits, $\bm{Z}_1 = \bm{Z}_2 = \bm{0}$ (defined below) and $\gamma\to \infty$ respectively.

After carefully repeating the derivation from Section \ref{sec:Towards A DDFT With External Flow} onwards, we use the Enskog approximation to resolve $f^{(2)}$ [see Eq.~\eqref{eq:reduced_phase_space_densities}]
\begin{align}
f^{(2)}(\vec{r}_1,\vec{r}_2,\vec{p}_1,\vec{p}_2,t) = f^{(1)}(\vec{r}_1,\vec{p}_1,t)f^{(1)}(\vec{r}_2,\vec{p}_2,t)g(\vec{r}_1,\vec{r}_2,[\varrho]),
\end{align}
where $g$ is the pair correlation function, which must be supplied by auxiliary means. Additionally we restrict the HIs to be two-body, thus
\begin{align}
\tilde{\bm{\Gamma}}_{ij}(\vec{r}^N) = \delta_{ij}\sum_{l\neq i}\bm{Z}_{1}(\vec{r}_i,\vec{r}_l)+(1-\delta_{ij})\bm{Z}_{2}(\vec{r}_i,\vec{r}_j),
\end{align}
where $\bm{Z}_1$ and $\bm{Z}_2$ are the diagonal and off diagonal blocks respectively of the translational component of the grand resistance matrix originating in the classical theory of low Reynolds' number hydrodynamics between suspended particles (see, e.g., \citet{jeffrey1984calculation}, \citet{happel2012low}). The corresponding DDFT is then
\begin{align}
\begin{cases}
&  \qquad\quad \partial_t\varrho(\vec{r}_1,t)+\nabla_{\vec{r}_1}\cdot\left(\varrho(\vec{r}_1,t)\vec{v}(\vec{r}_1,t)\right) =0,\\[2mm]
& \partial_t\vec{v}(\vec{r}_1,t)+(\vec{v}(\vec{r}_1,t)\cdot\nabla_{\vec{r}_1})\vec{v}(\vec{r}_1,t)\\
&\quad +\tfrac{1}{m}\nabla_{\vec{r}_1}\frac{\delta \mathcal{F}[\varrho]}{\delta \varrho}(\vec{r}_1,t)+ \gamma\vec{w}(\vec{r}_1,t)\\
&\quad+\gamma\int\mathrm{d}\vec{r}_2\,\Big[ \bm{Z}_1(\vec{r}_1,\vec{r}_2)\vec{w}(\vec{r}_1,t)+\bm{Z}_2(\vec{r}_1,\vec{r}_2)\vec{w}(\vec{r}_2,t)
)\Big]\\
&\qquad \qquad \times\varrho(\vec{r}_2,t)g(\vec{r}_1,\vec{r}_2,[\varrho]) = 0.
\end{cases}\label{eq:DDFT_HI}
\end{align}
We note that by setting the solvent velocity to zero we obtain the previously--derived DDFT including HIs in the case of a static background bath\cite{goddard2012unification}. In addition, setting the HI tensors to zero ($\bm{Z}_1 = \bm{Z}_2 = \bm{0}$) recovers the results of Ref.~\onlinecite{archer2009dynamical}.

\subsection{Overdamped Limit}

We now consider the overdamped limit with the inclusion of HI. Assuming that inter-particle HI are weak, that is $\bm{Z}_2\approx 0$, we obtain
\begin{align}\label{eq:no_acc_eqn}
&\tfrac{1}{m}\nabla_{\vec{r}_1}\tfrac{\delta \mathcal{F}[\varrho]}{\delta \varrho}(\vec{r}_1,t)+\gamma\bm{D}^{-1}(\vec{r}_1,[\varrho],t)(\vec{v}(\vec{r}_1,t)-\vec{u}(\vec{r}_1,t))\approx 0.
\end{align}
where $\bm{D}$ is the diffusion tensor, known rigorously to be positive definite (see, e.g., Ref.~\onlinecite{goddard2012overdamped}) given by 
\begin{align}
\bm{D}(\vec{r}_1,[\varrho],t):= \left[{\mathbf 1}+ \int\mathrm{d}\vec{r}_2\, \bm{Z}_1(\vec{r}_1,\vec{r}_2)\varrho(\vec{r}_2,t)g(\vec{r}_1,\vec{r}_2,[\varrho]\right]^{-1}  .
\end{align}
Now by adding and subtracting the same term to the continuity equation
\begin{align}
\partial_t\varrho + \nabla_{\vec{r}_1}\cdot \left(\varrho(\vec{v}-\vec{u})+ \varrho\vec{u}\right)  =0,\label{eq:cons_mass_incl_u}
\end{align}
and combining \eqref{eq:no_acc_eqn} and \eqref{eq:cons_mass_incl_u} we obtain
\begin{align}
&\partial_t\varrho(\vec{r}_1,t)+\nabla_{\vec{r}_1}\cdot\left(\varrho(\vec{r}_1,t)\vec{u}\right) \nonumber\\
&\quad\approx\tfrac{1}{m\gamma}\nabla_{\vec{r}_1}\cdot\left(\varrho(\vec{r}_1,t)\bm{D}(\vec{r}_1,[\varrho], t)\nabla_{\vec{r}_1}\tfrac{\delta \mathcal{F}[\varrho]}{\delta \varrho}\right), \label{eq:smoluchowski_eqn}
\end{align}
which may be compared with the DDFT in Eq.~(27) of Ref.~\onlinecite{RauscherDominguezKrugerPenna07}. Equation \eqref{eq:smoluchowski_eqn} is the Smoluchowski equation, differing from the one found in \citet{RauscherDominguezKrugerPenna07} and other approaches that start from overdamped Langevin dynamics.  We note that previous work has demonstrated that such differences arise even in the case when there is no background flow~\cite{goddard2012overdamped}.

\section{Results From Numerical Solution}\label{sec:numerical_solutions}

 \begin{figure*}[ht!]
  \centering

\includegraphics[width=\columnwidth]{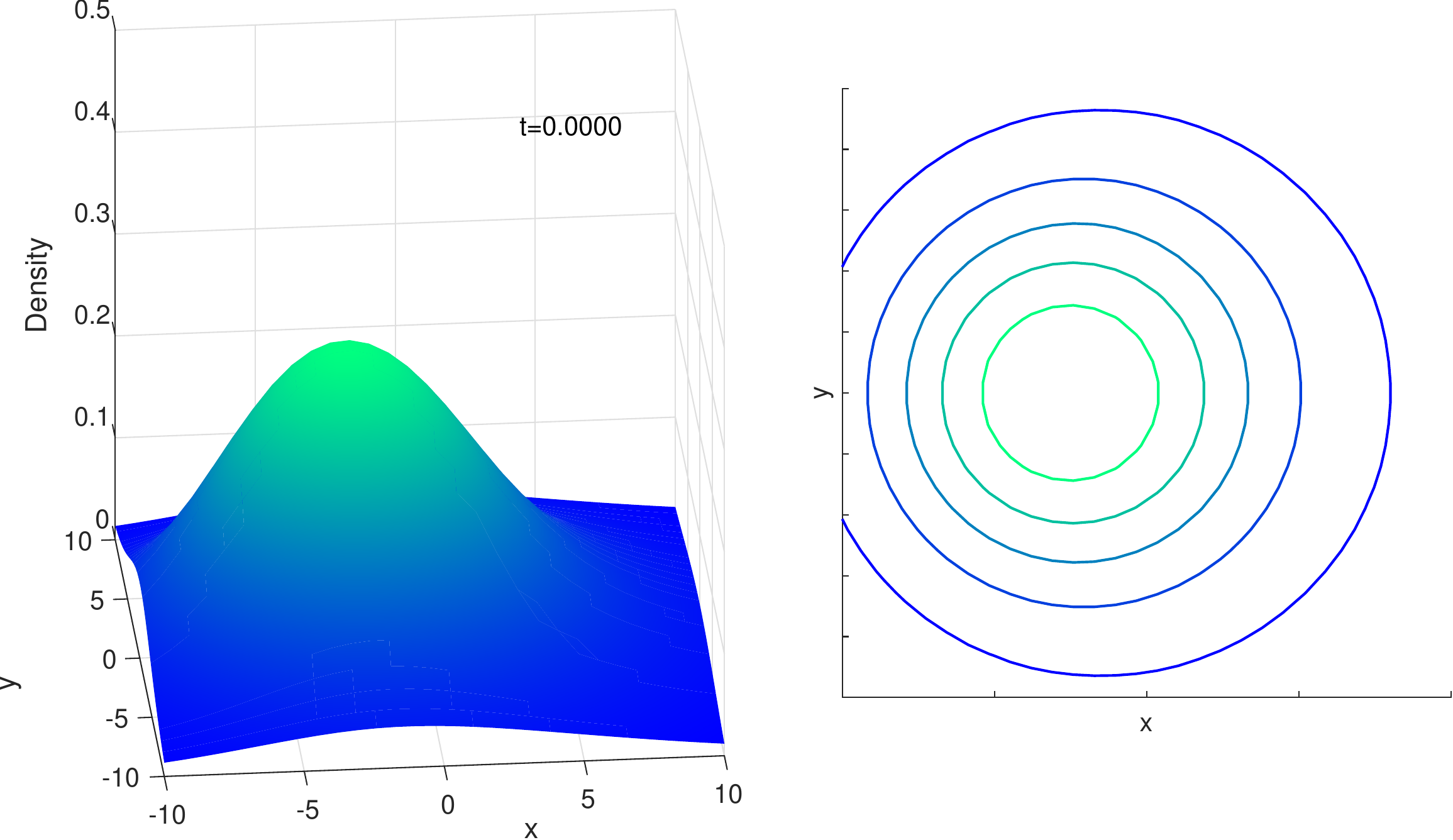}
\includegraphics[width=\columnwidth]{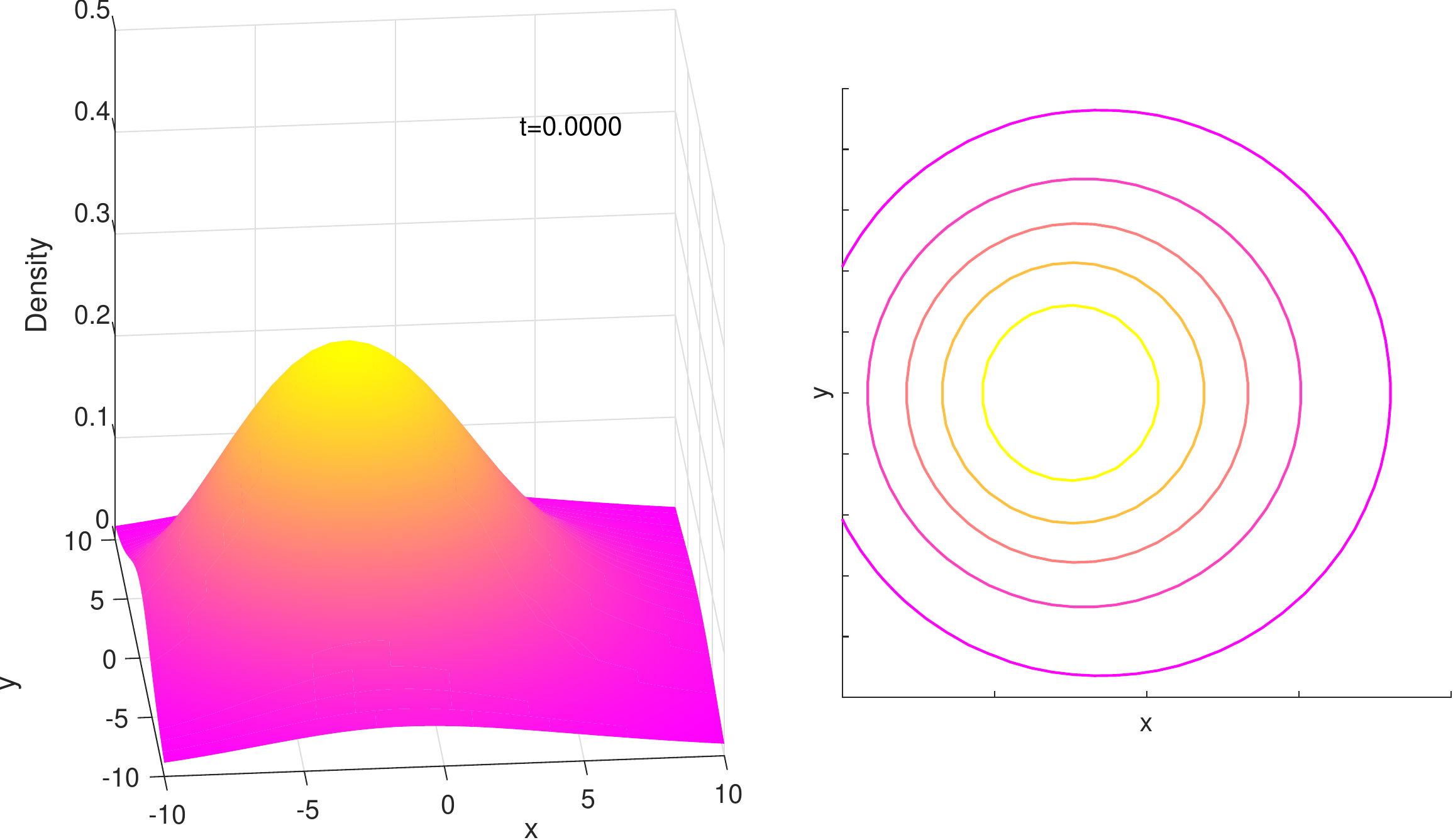}

\includegraphics[width=\columnwidth]{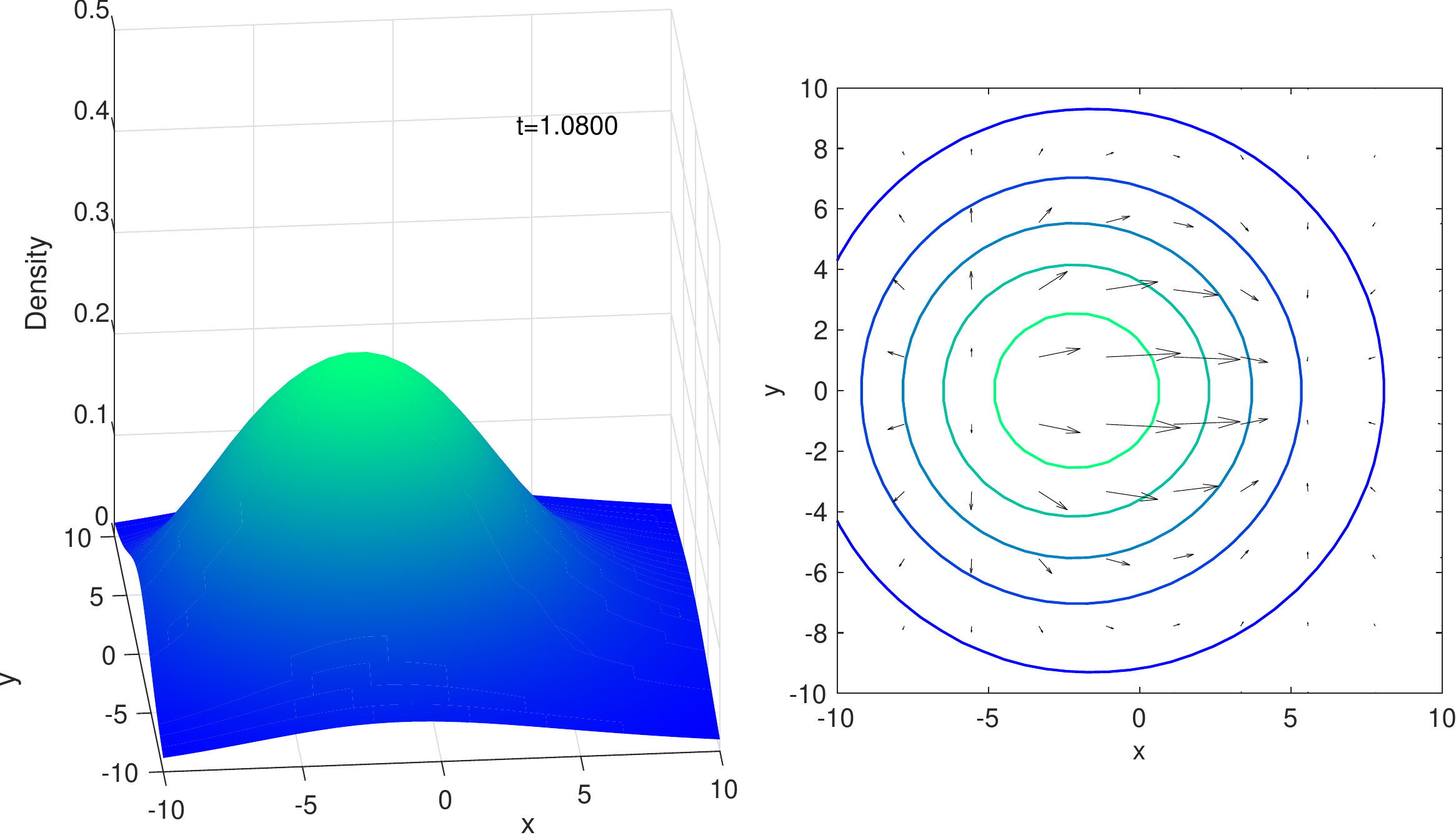}
\includegraphics[width=\columnwidth]{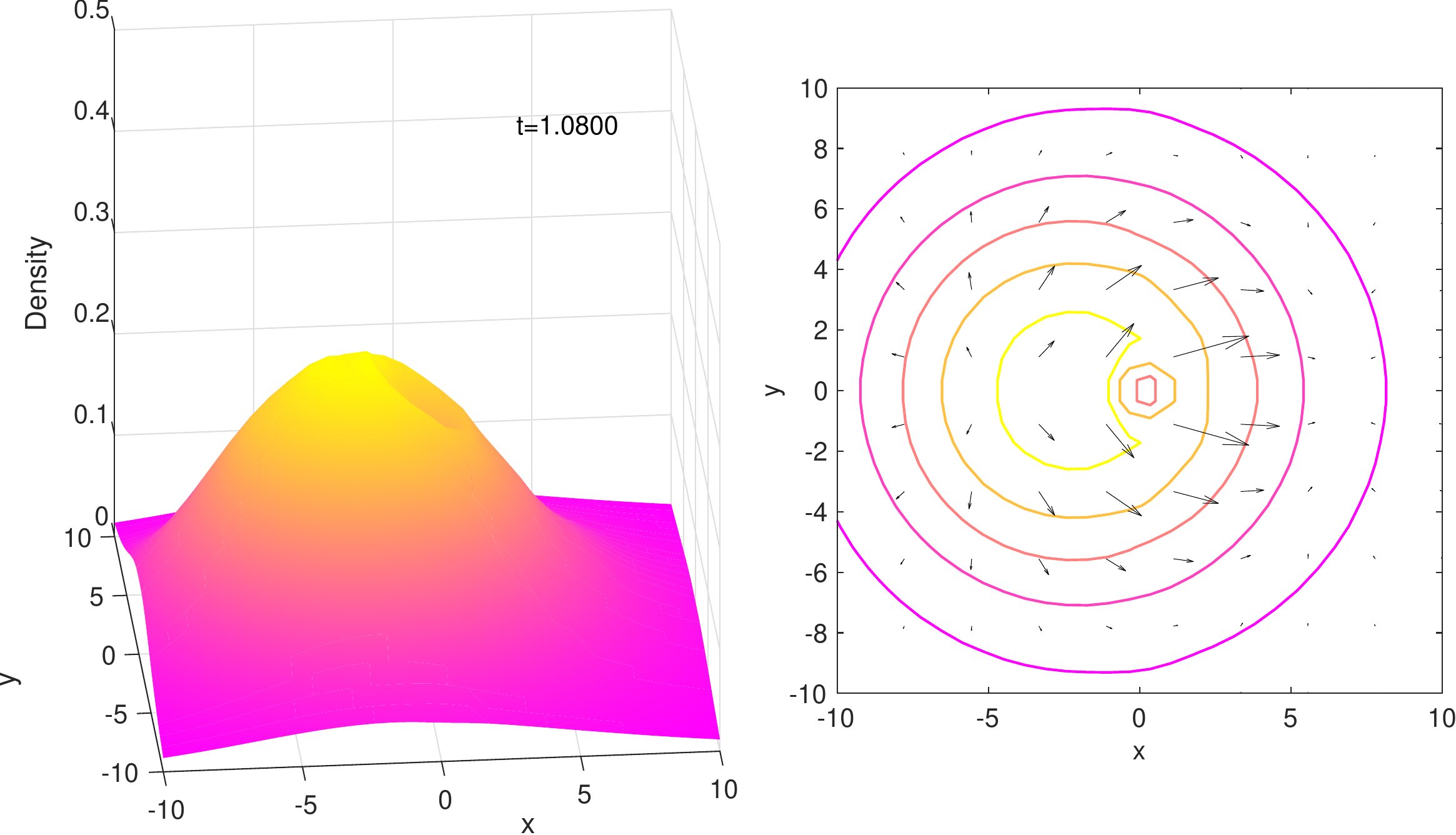}

\includegraphics[width=\columnwidth]{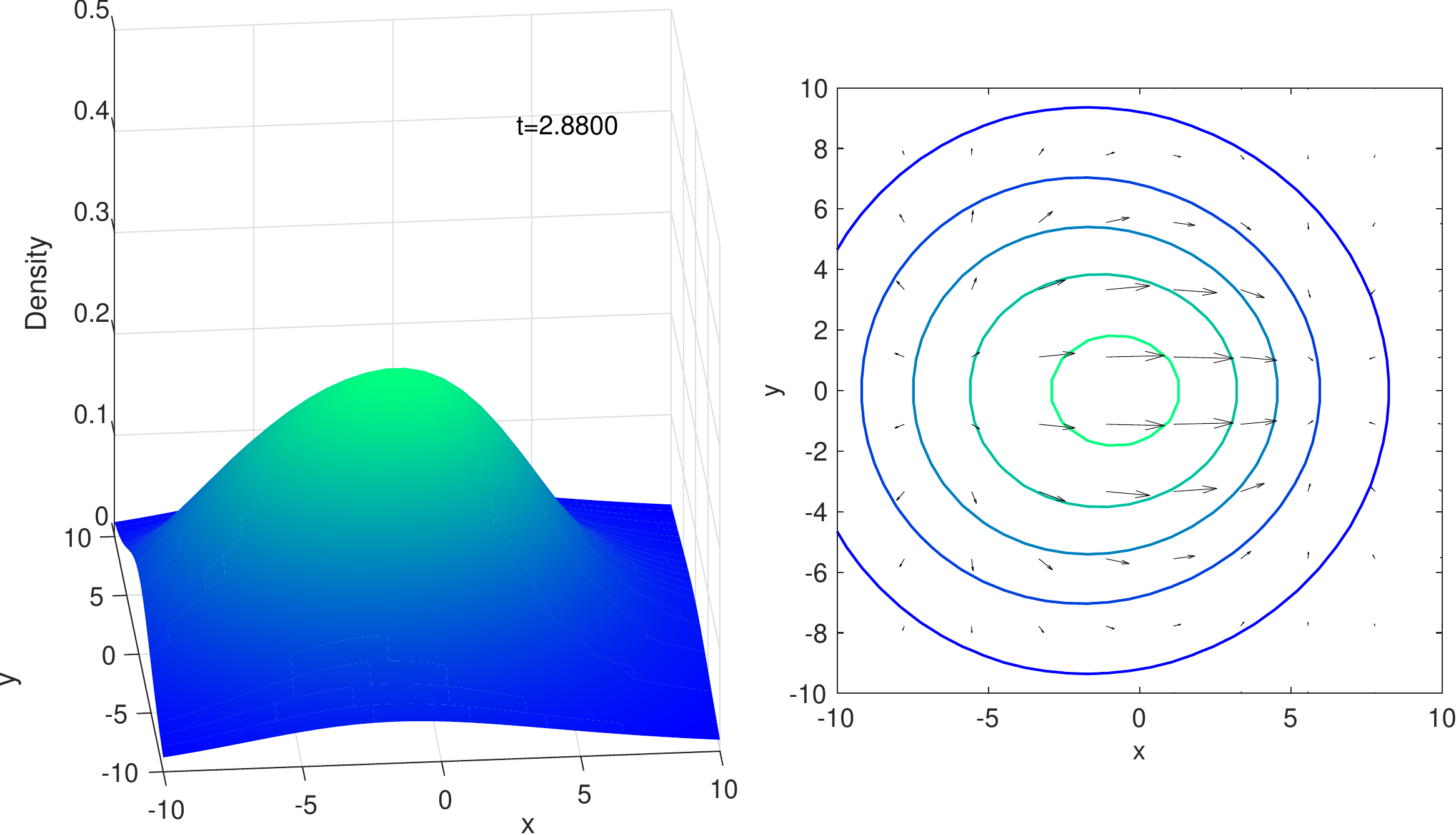}
\includegraphics[width=\columnwidth]{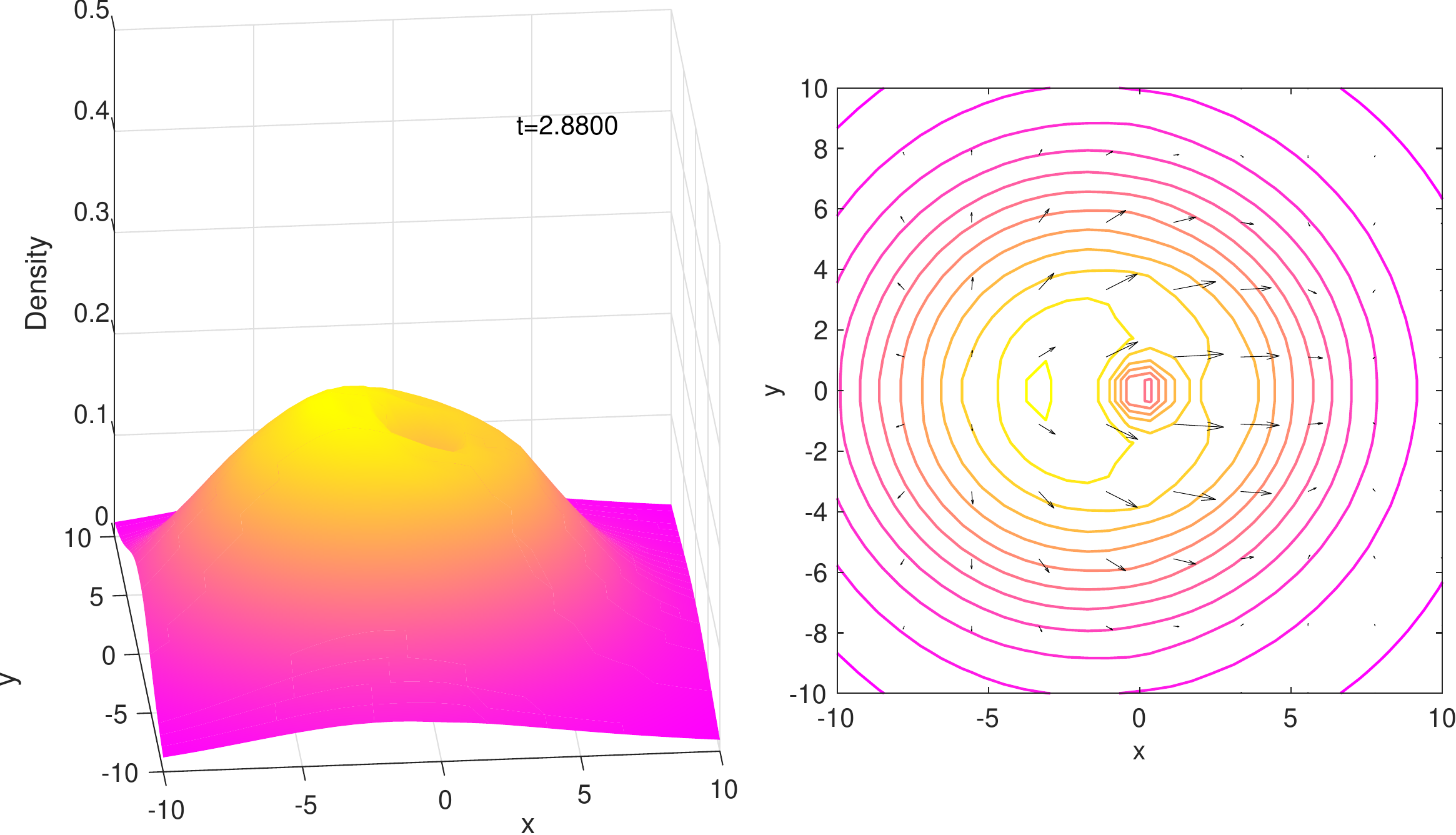}

\includegraphics[width=\columnwidth]{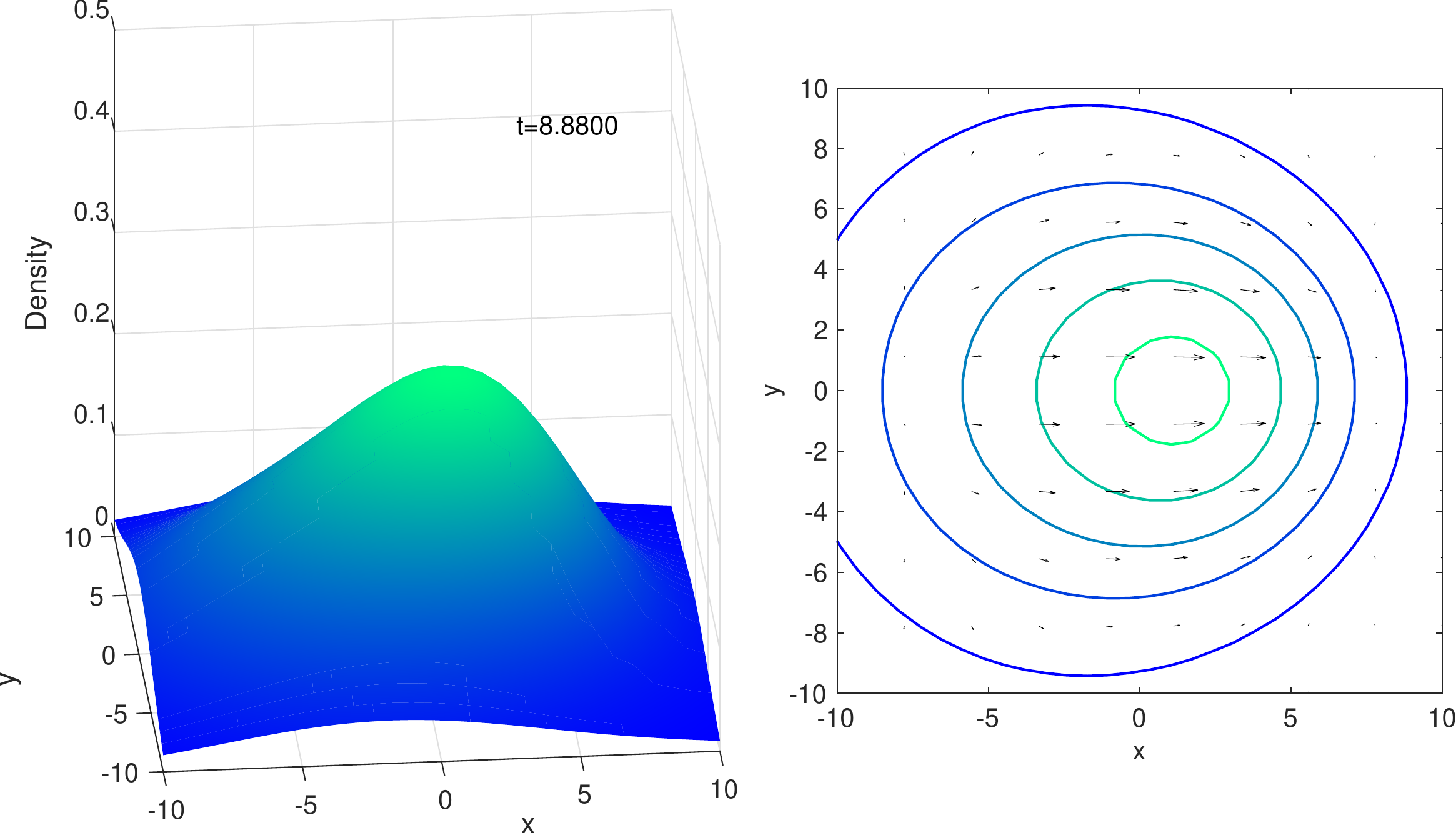}
\includegraphics[width=\columnwidth]{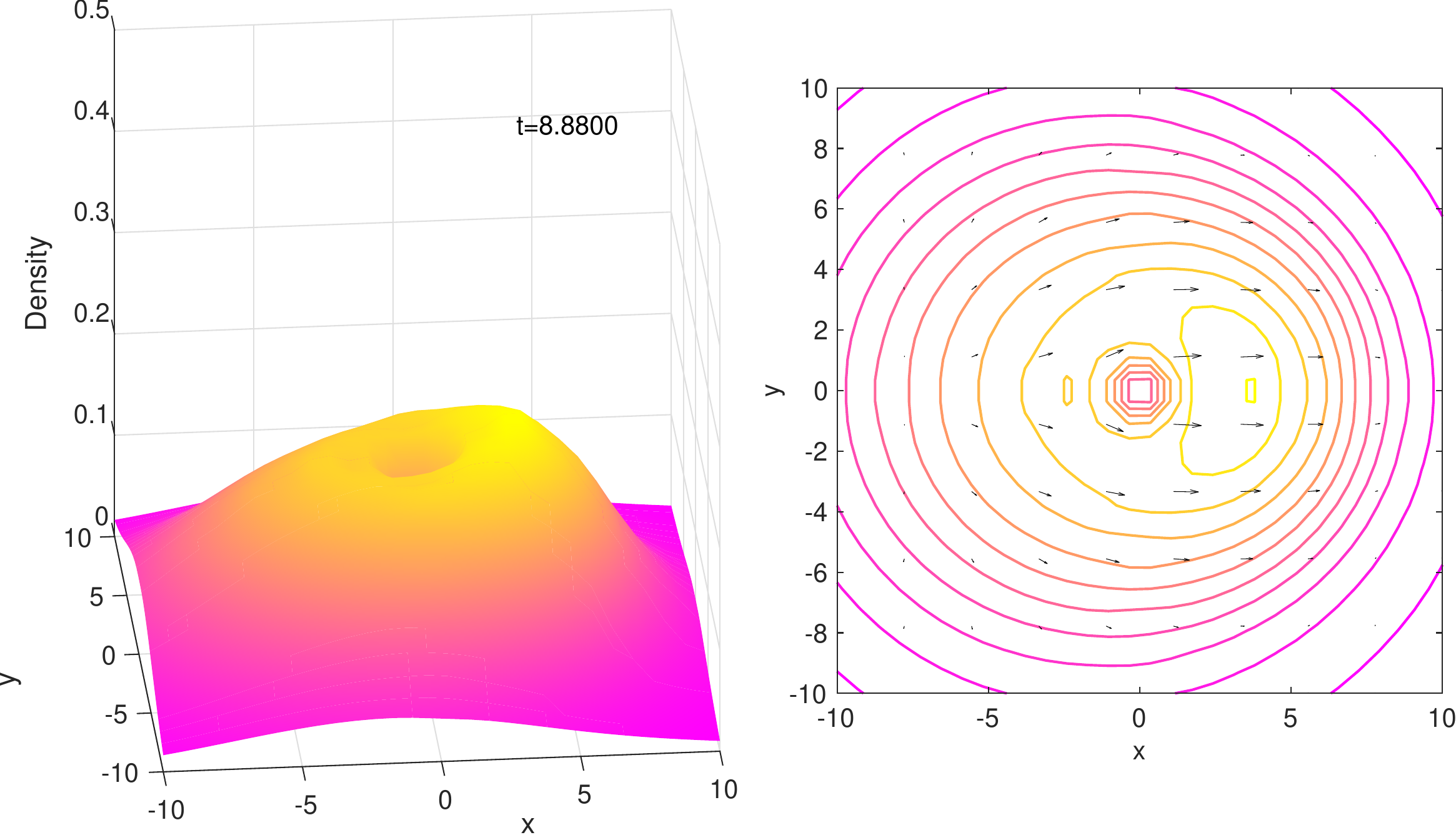}

\caption{Time snapshots of the density profile predicted by the DDFT \eqref{eq:DDFT}, with time increasing from top to bottom, as indicated on each subplot, and external potential given in Eq.~\eqref{eq:V1_t}, with $(x_a,\,y_a) = (-3,\,0) $ and $(x_b,\,y_b) = (3,\,0)$. In each case, we display both a surface plot and a contour plot, with the local flux vector $\vec{j}$ shown by black arrows displayed on top of the contour plots. The four plots on the left are for the case of zero background flow $\vec{u} = 0$, and the four on the right are for a line source, i.e., uniform flow with $\vec{u}$ given by Eqs.~\eqref{eq:line_source_u}--\eqref{eq:line_source_v}.}\label{fig:line_source_flow}
\end{figure*}

The aim of this section is to illustrate that the inclusion of general, incompressible background flows in our theory is a non-trivial addition to DDFT, and can, in fact, lead to interesting differences in the dynamics of the density and velocity of colloidal dispersions when compared to non-driven systems. We also wish to demonstrate that the DDFTs described above are numerically tractable, at least in two spatial dimensions.  We do this by presenting a selection of numerical computations of the derived DDFT \eqref{eq:DDFT} for a system of hard spheres with excess Helmholtz free energy functional $\mathcal{F}_{ex}[\varrho]$ given by fundamental measure theory (FMT), and compare the resulting dynamics to previously derived DDFTs, specifically the inertial DDFT of Ref.~\onlinecite{archer2009dynamical} and the overdamped with external flow theory of Ref.~\onlinecite{RauscherDominguezKrugerPenna07}. The choice of excess free energy functional to model hard spheres is not restrictive, and, in particular, our DDFT can also be used to describe the flow of other systems through the choice of a different approximation for $\mathcal{F}_{ex}[\varrho]$. For example, we have also tested our DDFT in conjunction with a simple mean-field approximation for $\mathcal{F}_{ex}[\varrho]$ relevant to soft-core particles, but do not present the results here.

All of the numerical simulations presented here are performed using pseudospectral methods, available in the MATLAB 2DChebClass package~\cite{DDFTCode}.

The first example we present concerns a solvent flow taking the form of a line source in a uniform field that is inspired by two-dimensional irrotational flow from classical fluid mechanics. The second example considers an external flow field with finite vorticity and a local sink, thereby demonstrating the general validity of the presented DDFT when subject to flows with a rotational component as well as the effects of inertia. Such rotational flows can be considered as a model for forced-flow layer liquid chromatography \cite{sherma2003handbook}, whereas the addition of a local sink models the flow of colloids around a plug-hole-like exit. The third example concerns flow in a confined geometry (a periodic two-dimensional capillary) for a simple model system in haemodynamics \cite{yu2017microstructure}. Here we demonstrate the effect of inertia in response to an instantaneous change in the direction of the applied background flow, showing agreement with Ref.~\onlinecite{RauscherDominguezKrugerPenna07} both in the high friction limit, and in equilibrium.

For all our numerical experiments we non-dimensionalize the equations with the units of length, mass and energy being $\varsigma$ (colloid diameter), $m$ (colloid mass) and $k_B T$ (Boltzmann's constant times the temperature), respectively. For examples in unbounded domains, we also artificially modify the solvent velocity in the far field, in order to obtain finite solutions to our computations. This is necessary when using a pseudo-spectral collocation scheme, where on unconfined domains there exist computational points at infinity. This is only done in regions of the computational domain where the density $\rho(\vec{r},t)$ is infinitesimal or zero (in particular, at infinity) and is achieved by smoothly reducing the solvent velocity to zero away from regions with finite density (e.g., through multiplication by an appropriately--chosen error function).  The results of the simulations are insensitive to the exact choice of this methodology.

We choose the colloidal particles to be hard spheres, with pair potential $V_2(|\vec{r}-\vec{r}'|) = \infty$ for $|\vec{r}-\vec{r}'| < \varsigma$ and zero otherwise, where $\vec{r}$ and $\vec{r}'$ denote the positions of two different colloidal particles. This hard sphere exclusion is described by the excess free energy $\mathcal{F}_{ex}[\varrho]$; here we use the three-dimensional FMT functional of \citet{rosenfeld1989free}, averaged into two dimensions. This means that though our numerical solutions are restricted to two dimensions, the physical modelling of the hard spheres is maintained in three dimensions. Additionally, we do not consider any \citet{faxen1922widerstand} type corrections to the solvent fluid (c.f., \citet{RauscherDominguezKrugerPenna07,Rauscher10}) whereby we mean that, in reality, the colloidal particle geometry is known to perturb the flowing solvent field local to each particle. In effect, we assume that the solvent flows through the particles as it carries them in the flow. We also assume only moderately low density particle systems, due to the known deficiencies in the local steady approximation to accurately predict the dynamics far from steady state \cite{almenar2011dynamics}. 
 
We remark now on initial conditions: In each of the numerical examples presented we assume $\vec{u} = 0$ at $t=0$, but thereafter ($t>0$) $\vec{u}$ instantaneously changes to a non-zero flow. The reason for this is twofold: (i) it permits the existence of an initial density distribution compatible with the present system of PDEs \eqref{eq:DDFT}; (ii) it allows us to examine the transient behaviour between the different DDFTs without including dynamic effects arising due to different initial data. Obtaining the initial equilibrium density distribution involves solving the DFT\cite{evans1979nature,wu2007density} problem $(\delta \mathcal{F}[\varrho]/\delta \varrho) = 0$. We obtain the initial condition via Picard iteration, a standard technique within the DFT community, see e.g., \citet{roth2010fundamental}, and which can also be viewed as a damped version of standard fixed point algorithms. The numerical examples presented here were determined to be converged in the number of collocation points $M$. In particular, we repeated the computations under increasing $M$, until the time dependent dynamics appeared constant in $M$. 


\subsection{Line Source in a Uniform Flow}\label{num:line_source}

\begin{figure*}[ht!]
  \centering

\includegraphics[width=\columnwidth]{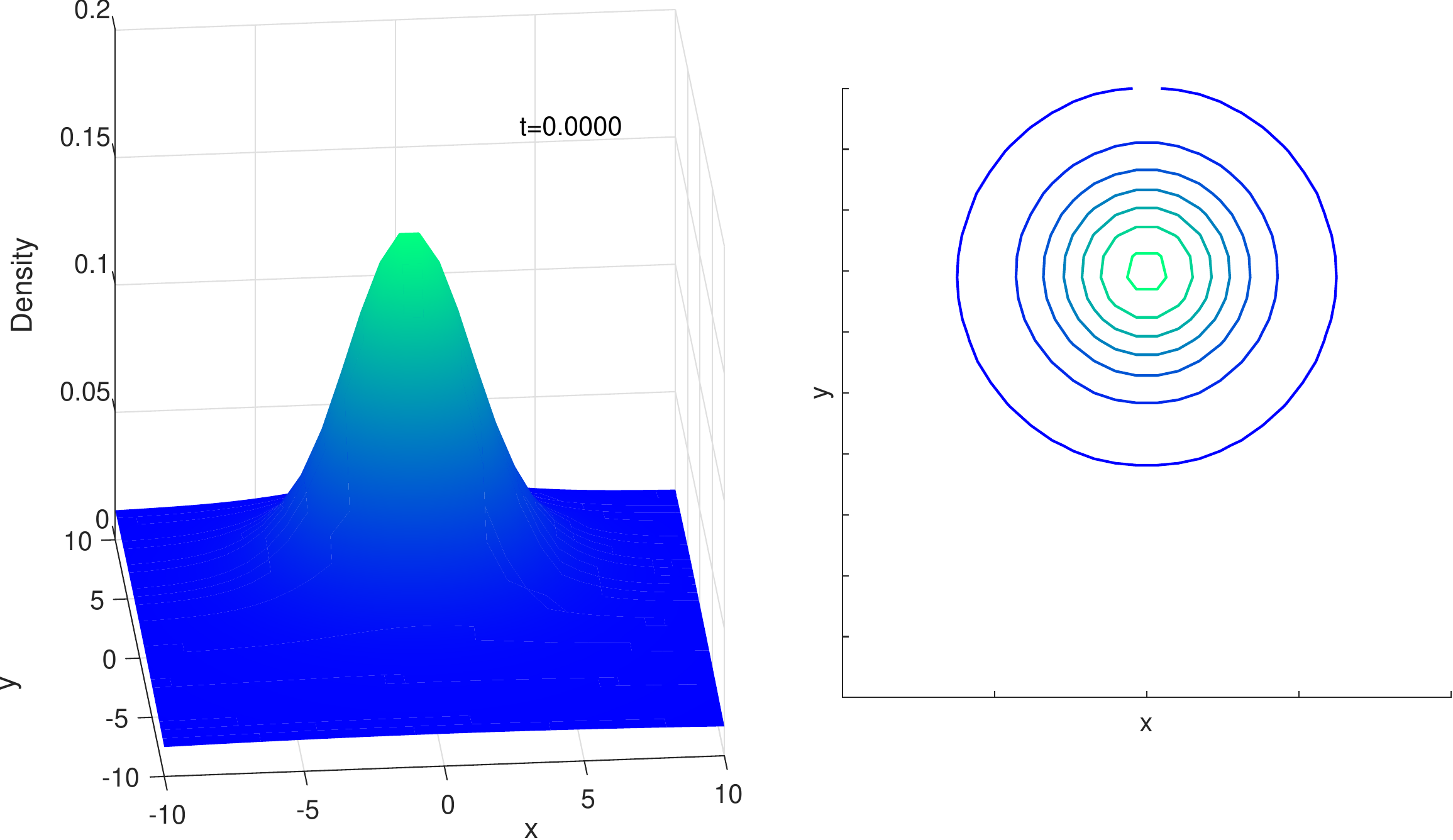}
\includegraphics[width=\columnwidth]{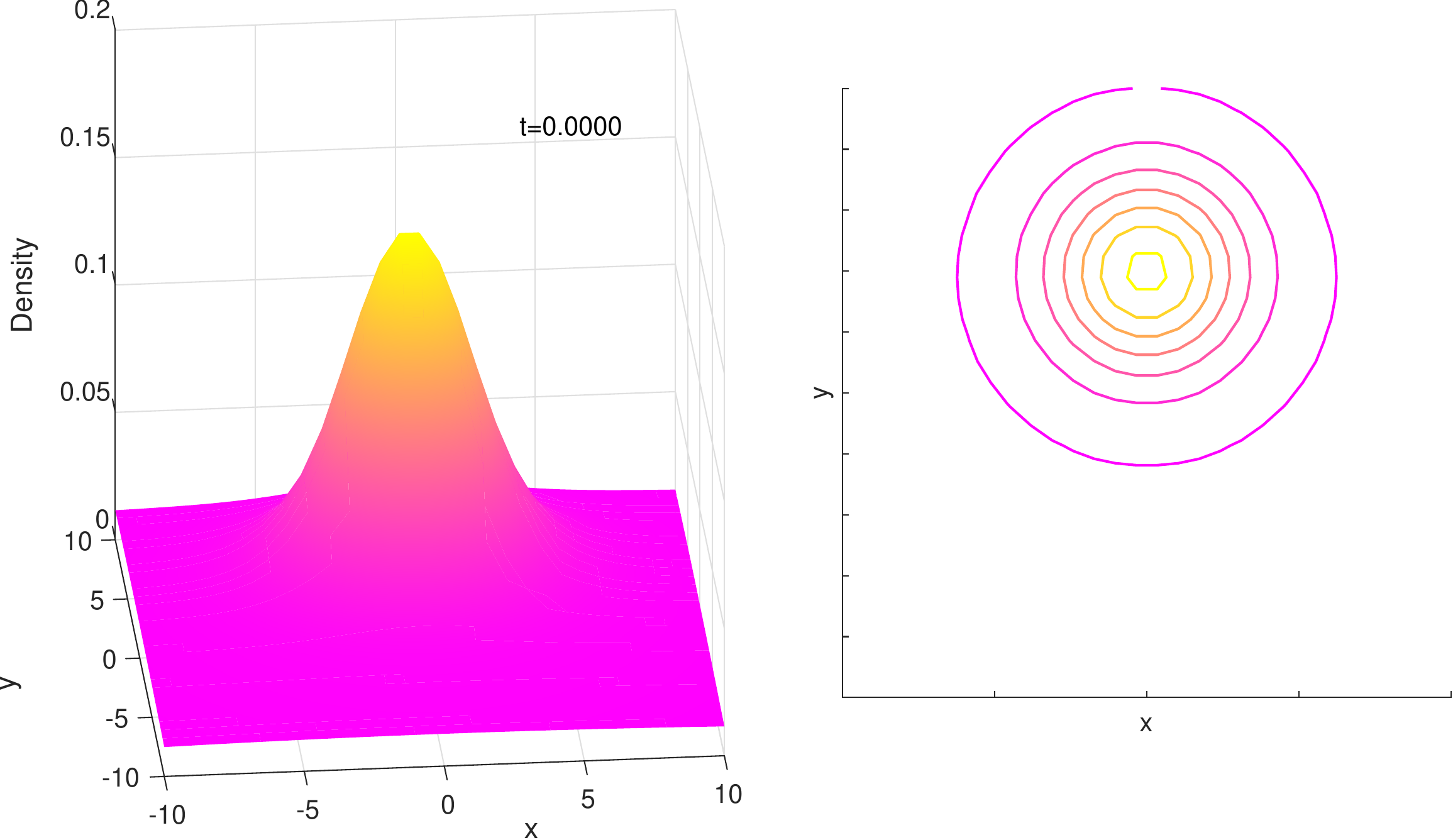}

\includegraphics[width=\columnwidth]{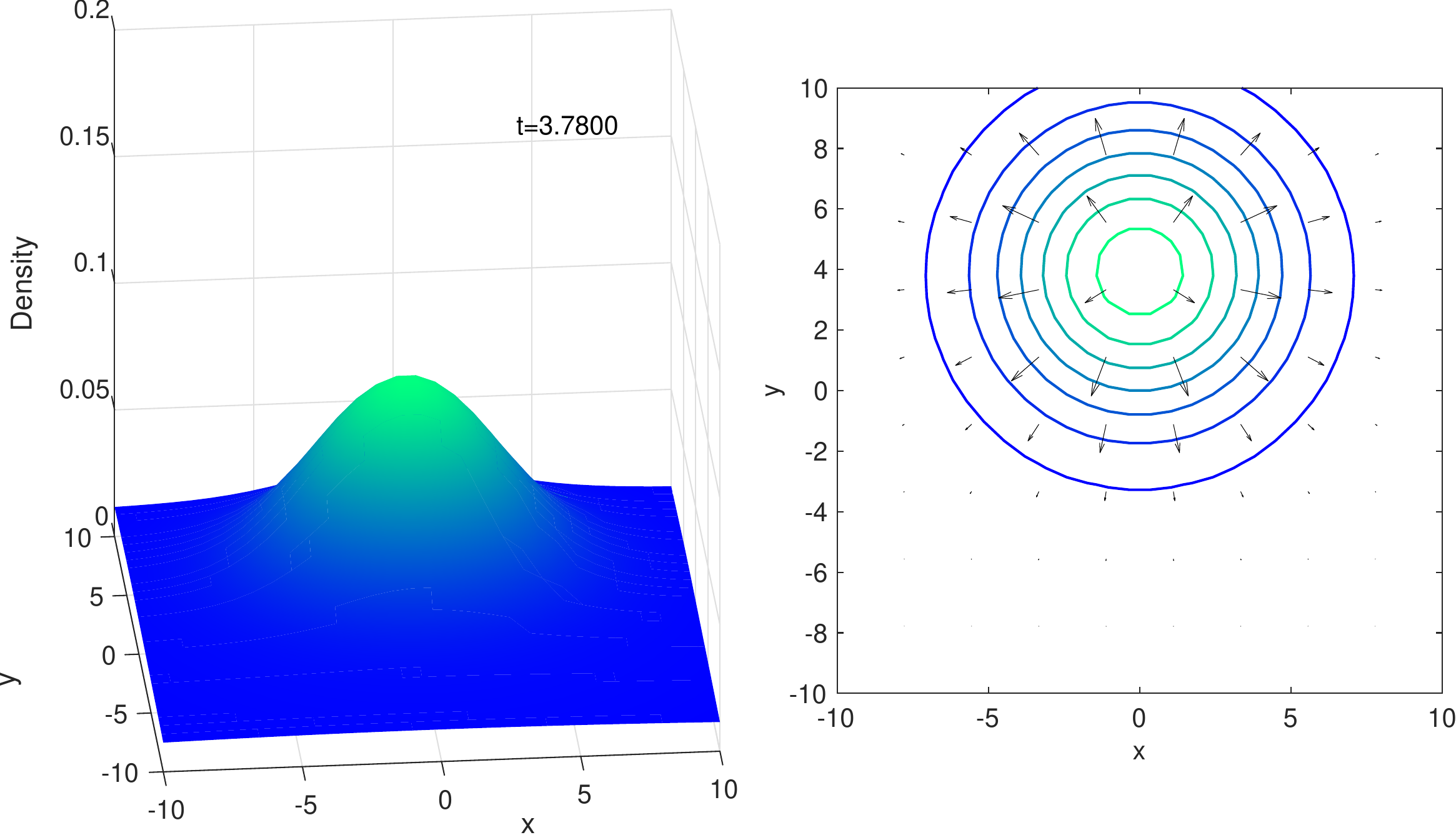}
\includegraphics[width=\columnwidth]{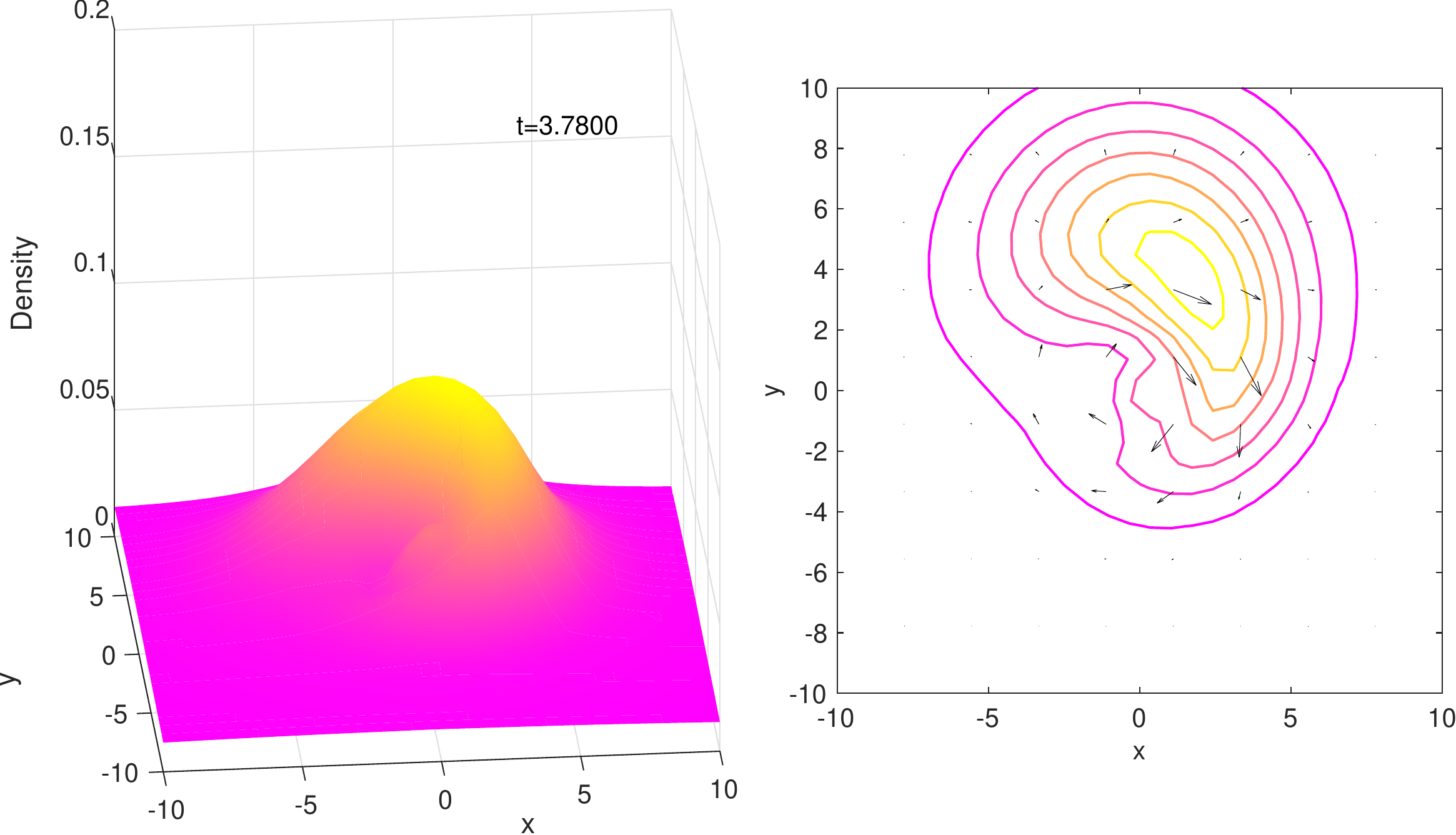}

\includegraphics[width=\columnwidth]{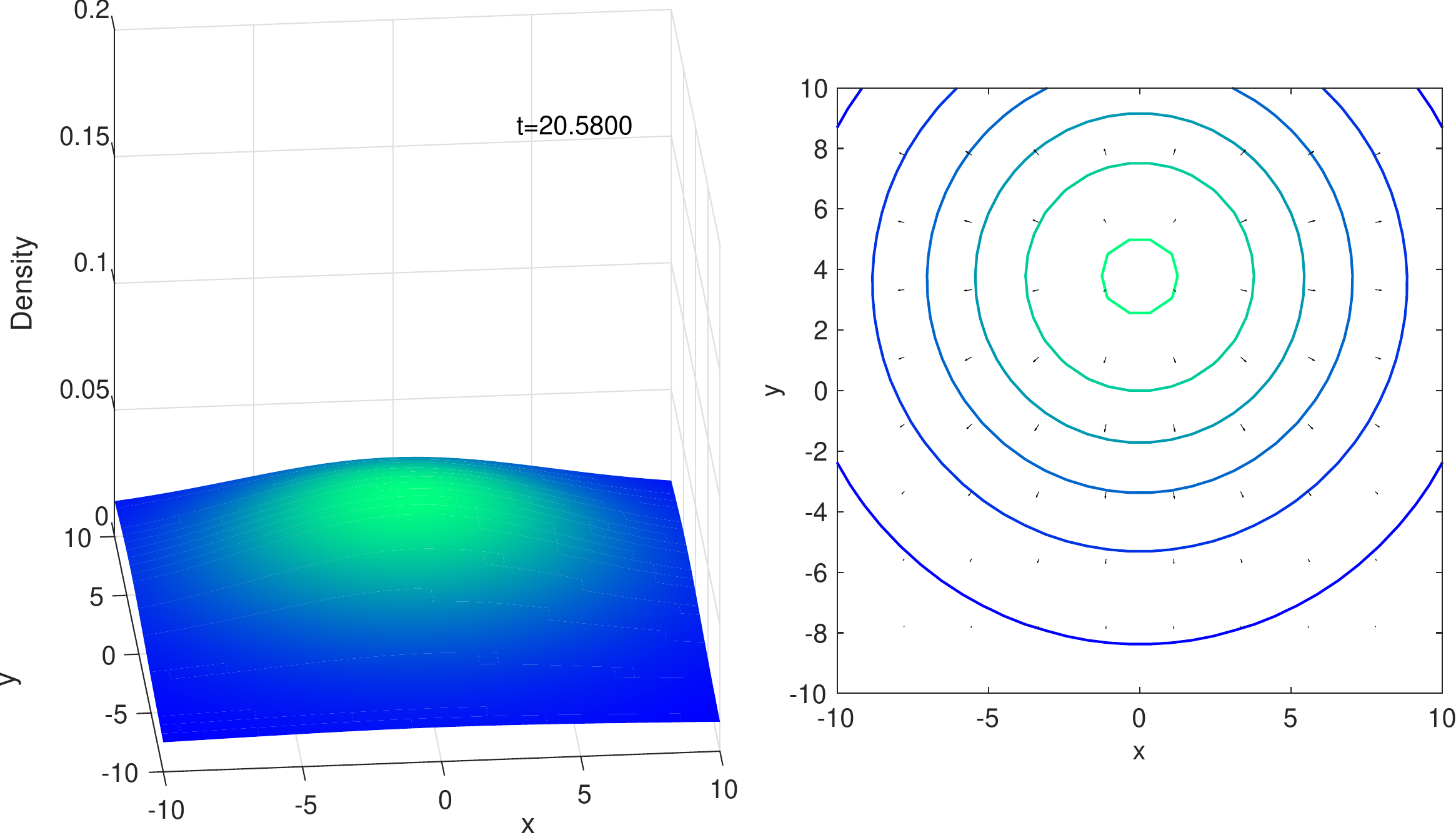}
\includegraphics[width=\columnwidth]{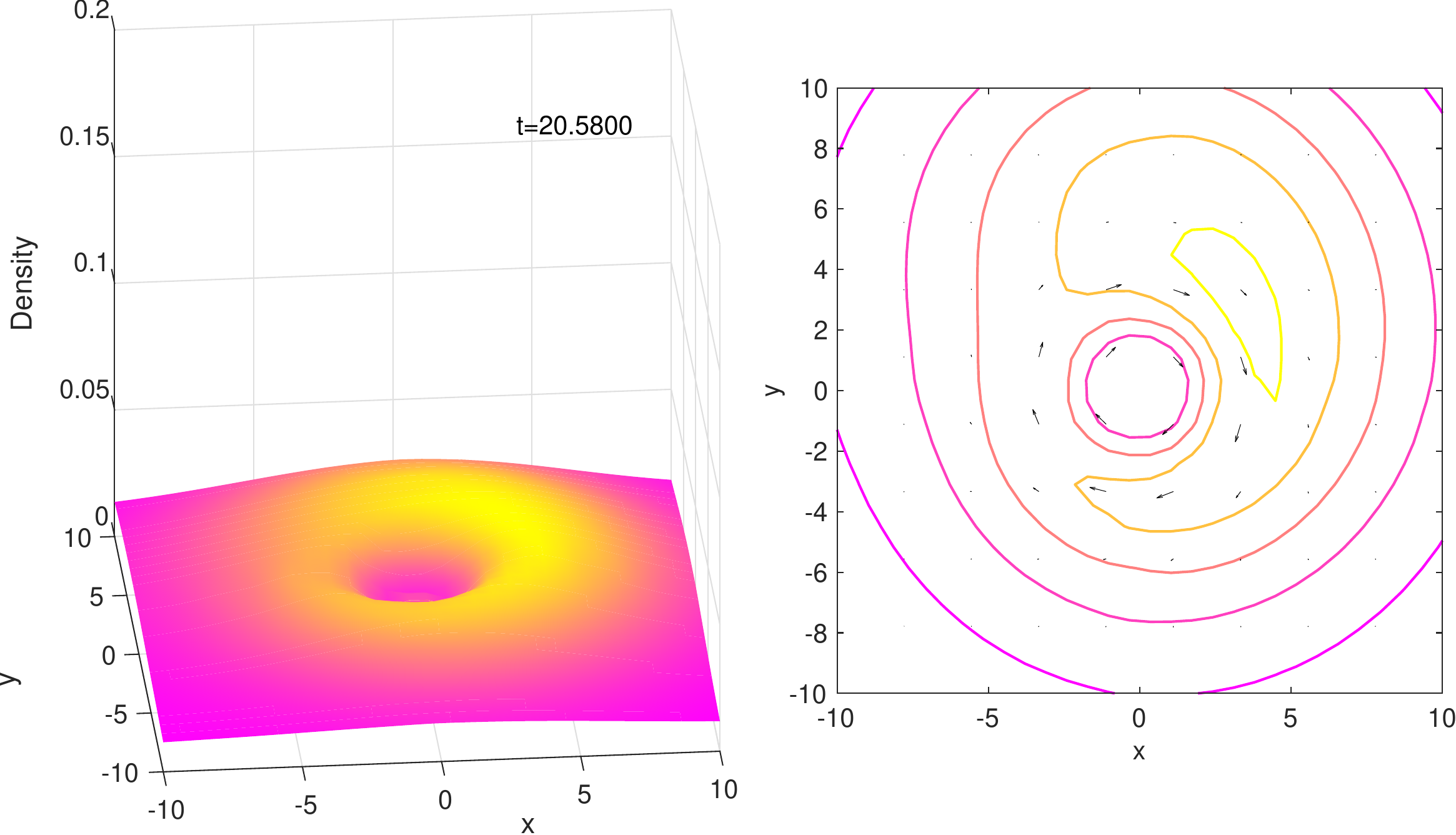}

\includegraphics[width=\columnwidth]{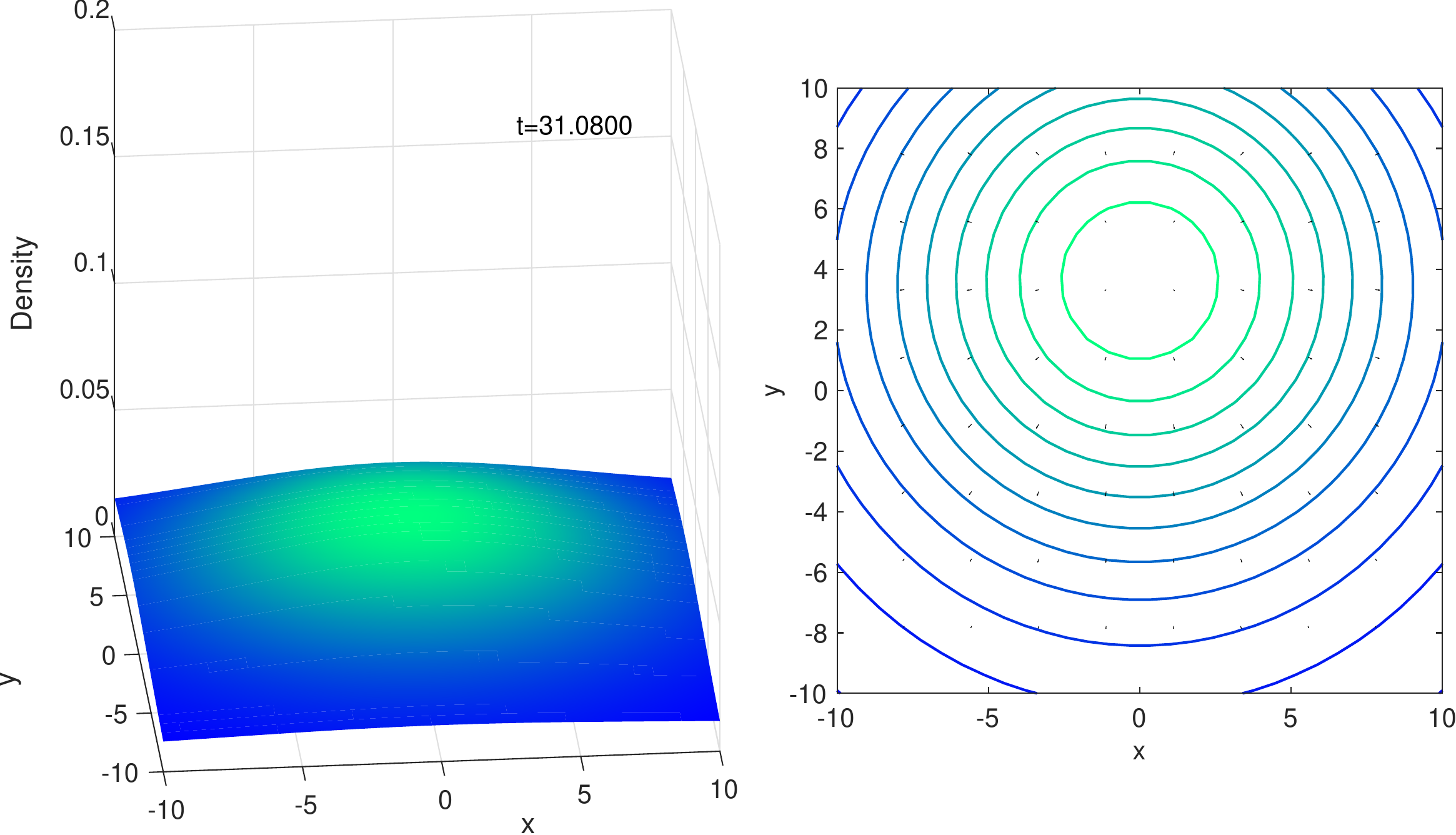}
\includegraphics[width=\columnwidth]{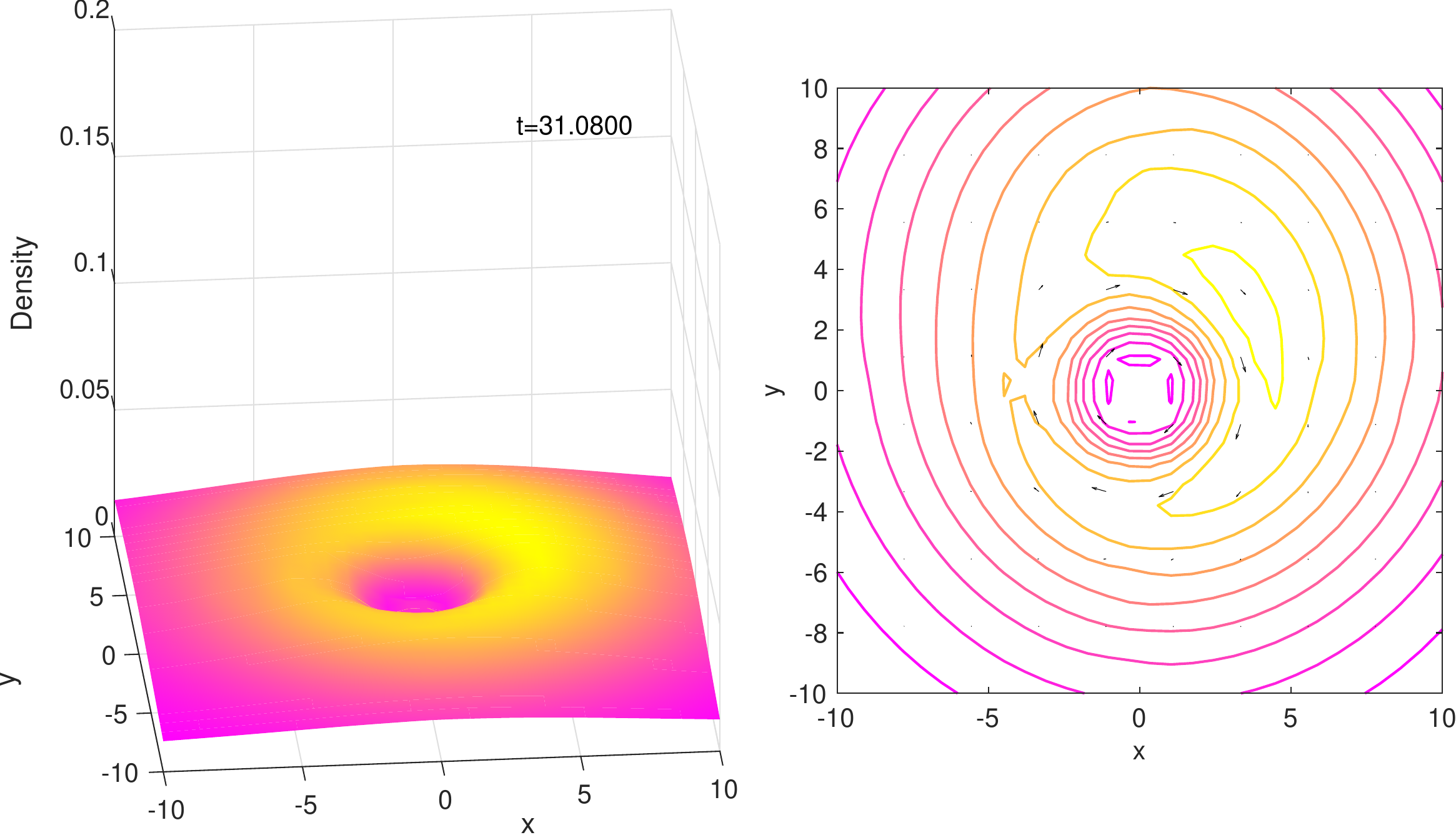}

\caption{Time snapshots of the density profile predicted by the DDFT \eqref{eq:DDFT}, with time increasing from top to bottom, as indicated on each subplot, and external potential given in Eq.~\eqref{eq:V1_fig2}, with $(x_a,\,y_a) = (0,\,4)$. In each case, we display both a surface plot and a contour plot, with the local flux vector $\vec{j}$ shown by black arrows displayed on top of the contour plots. The four plots on the left are for the case of zero background flow $\vec{u} = 0$, and the four on the right are for the case where $\vec{u}$ given by \eqref{eq:vortex_sink_u}--\eqref{eq:vortex_sink_v}.}
\label{fig:sink_flow}
\end{figure*}


This example is a model for colloidal particles subject to an optical trap in the presence of a background solvent flow emanating from a point source, and moving uniformly throughout the $x$-$y$ plane. The background flow is defined as follows: Let $\vec{e}_x$ and $\vec{e}_y$ denote the canonical basis vectors of the two-dimensional Euclidean plane in the $x$ and $y$ directions, respectively. We consider an external flow of the form $\vec{u}(x,y) = \left(u(x,y),\, v(x,y)\right)^\top$ where $u(x,y)$ and $v(x,y)$ are the scalar velocity fields of the solvent in the $\vec{e}_x$ and $\vec{e}_y$ directions respectively, given by 
\begin{subequations}
\begin{align}
u &= U + \frac{Q x}{2\pi (x^2 + y^2)},\label{eq:line_source_u}\\
v &= \frac{Q y}{2\pi (x^2 + y^2)}, \label{eq:line_source_v}
\end{align}
\end{subequations}
where $U$ and $Q$ are the magnitudes of the uniform and line flow sources, respectively. We note that $\vec{u}(x,y)$ is derived from the gradient of the potential $\phi(x,y) = U x + Q/(2\pi) \log \sqrt{x^2+y^2}$. There is a stagnation point (i.e.~point where the velocity is zero) in the flow, at $(x,\, y) = (-Q/(2\pi U),\, 0)$, and a separatrix (crossing the $x$-axis through this point) separates the fluid emanating from the source (at $(x,\, y) = (0,\, 0)$) from the fluid carried by the uniform flow. To illustrate the influence of this flow field, we choose $U = 0.1$ and $Q = 0.35$ and use a confining quadratic external potential $V_1(x,y) = \lambda(x^2+y^2)$ to force the particles to remain in the vicinity of the origin, a choice which has been shown to not induce layering (i.e.\ oscillations in the particle density profile due to packing against the boundary). For $t > 0$, the external potential $V_1$ is modified to 
\begin{align}
V_1(x,y,t)= &\lambda(x^2+y^2)\\
&-\lambda e^{-\left(\frac{t}{\tau}\right)^2}e^{-\left(\frac{(x-x_a)^2}{\sigma^2} + \frac{(y-y_a)^2}{\sigma^2}\right)} \\
&- \lambda\left(1-e^{-\left(\frac{t}{\tau}\right)^2}\right)
e^{-\left(\frac{(x-x_b)^2}{\sigma^2} +\frac{(y-y_b)^2}{\sigma^2}\right)}.
\label{eq:V1_t}
\end{align}
We choose $(x_a,\,y_a) = (-3,\,0) $ and $(x_b,\,y_b) = (3,\,0)$, thereby moving the minimum in $V_1$ from near $(x_a,\,y_a)$ to near $(x_b,\,y_b)$, which drives the density across the stagnation point. We also choose $\lambda = 0.01$, $\sigma = 50$, $\tau = 0.01$ and $\gamma = 2$ for $N = 50$ hard spheres. In Figure \ref{fig:line_source_flow} we show the transient densities $\varrho(\vec{r},t)$ and velocities $\vec{v}(\vec{r},t)$ obtained from \eqref{eq:DDFT}, alongside the solutions obtained with external flow turned off ($\vec{u} = 0$), for comparison. We see substantial differences between density solutions with and without external flow. We find that the chosen termination time for the dynamics, $t = 12$, is sufficient to observe both systems reach steady state (not shown). In the presence of external flow (yellow-pink density plots) we observe that the density peak is advected across the stagnation point at $(-Q/(2\pi U),\, 0)\approx(-0.56,\,0)$, which generates a dip in the density profile in the vicinity of that point.

\subsection{Flow in a Localised Vortex and Sink }

\begin{figure*}[th!]
\includegraphics[width = 0.8\textwidth]{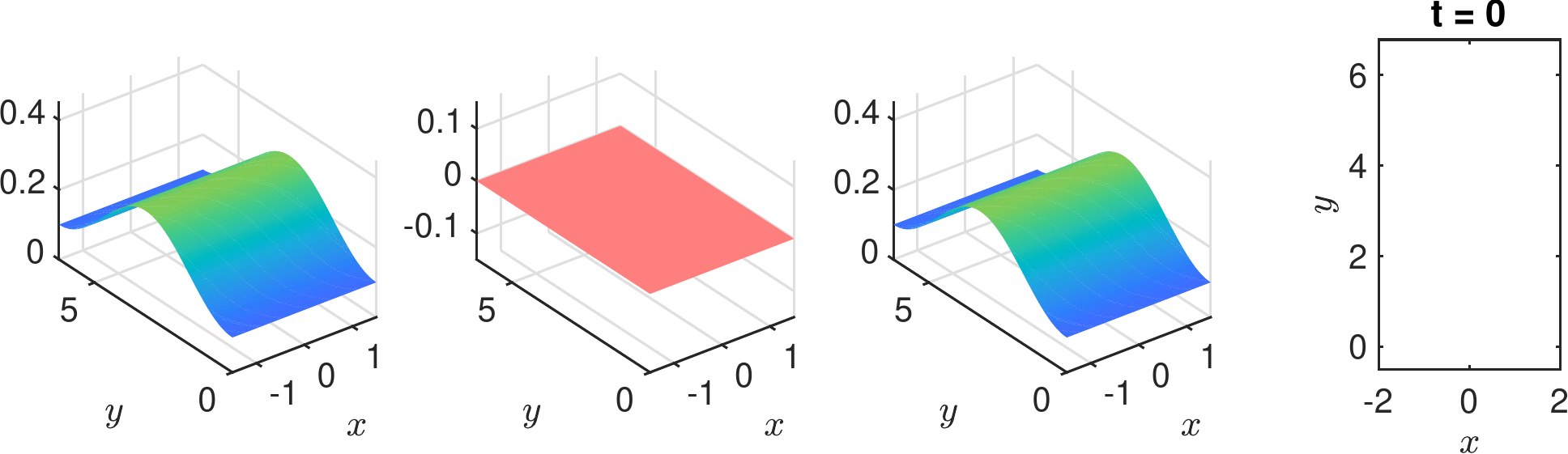}\\
\includegraphics[width = 0.8\textwidth]{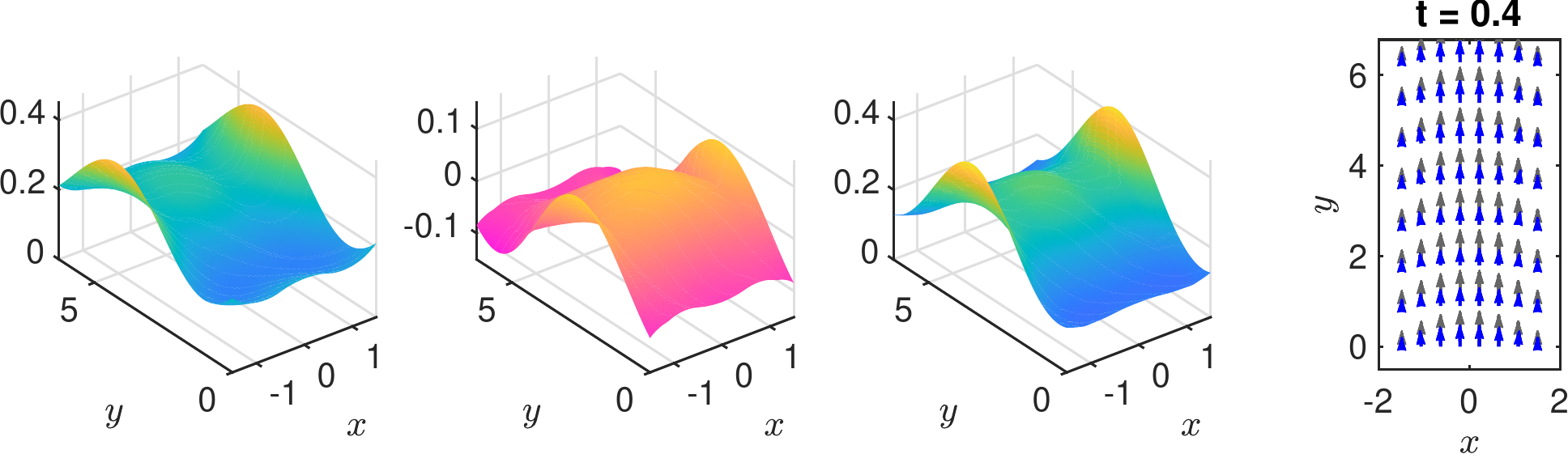}\\
\includegraphics[width = 0.8\textwidth]{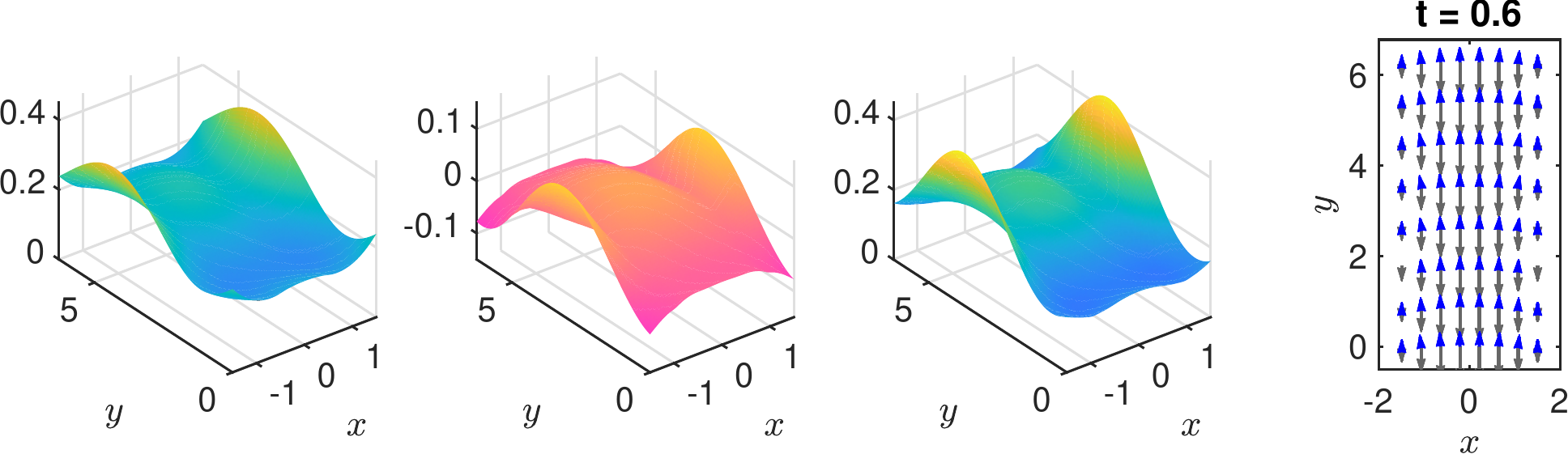}\\
\includegraphics[width = 0.8\textwidth]{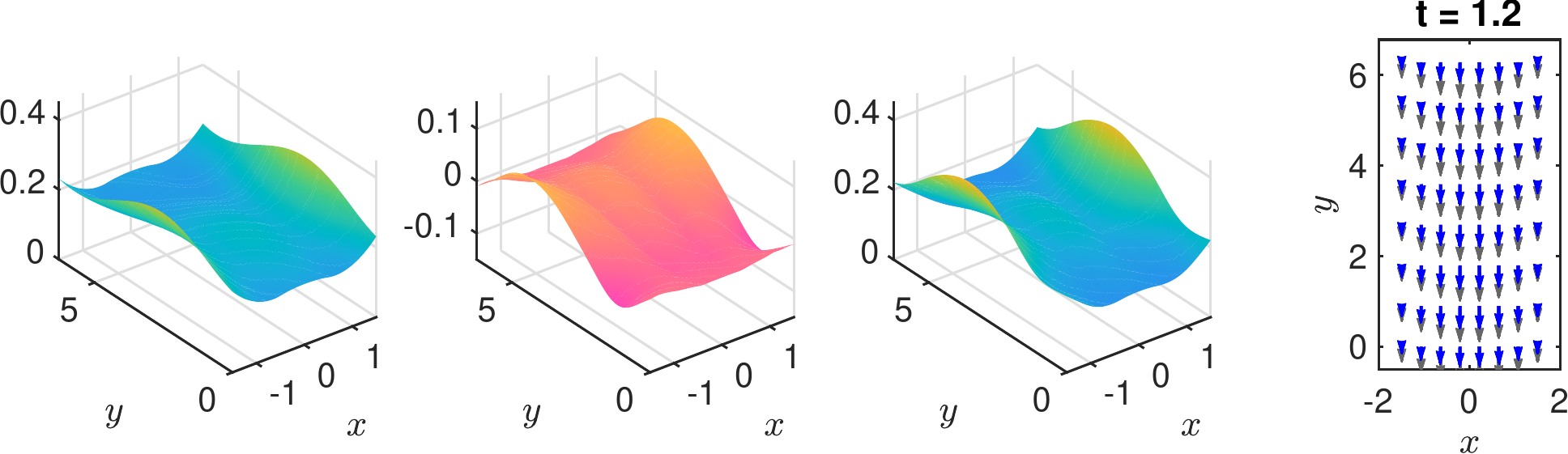}\\
\includegraphics[width = 0.8\textwidth]{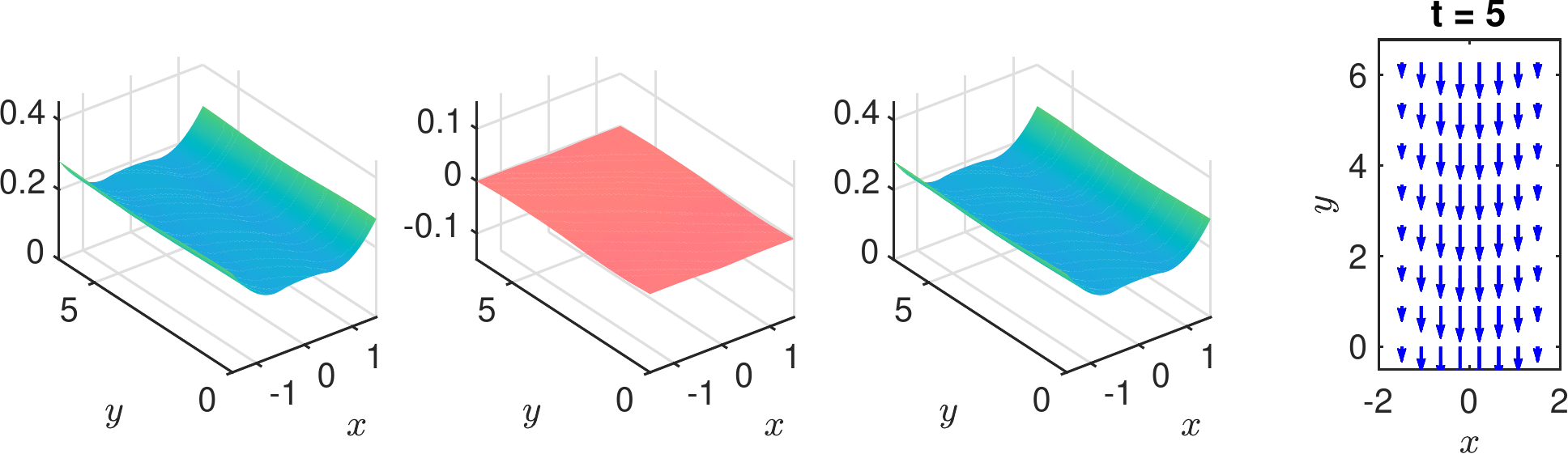}
\caption{DDFT results for the density and current for flow in a channel.  In each row the first and third plots show the densities resulting from overdamped and underdamped DDFT, respectively; the second plot shows the difference between the two densities ($\rho_o - \rho_u$); the fourth plot shows the velocity of the underdamped dynamics (blue arrows), and the background flow (grey arrows).  Each row corresponds to a different time: $t = 0, 0.4, 0.6, 1.2, 5$.}
\label{fig:channelFlow}
\end{figure*}

To more fully understand the effect of including an external flow under inertial dynamics, we now study a simple model including finite vorticity. We consider an external flow field of the form
\begin{subequations}
\begin{align}
u &= R(y-x_{r_0}) + \frac{Q x}{2\pi (x^2 + y^2)}, \label{eq:vortex_sink_u}\\
v &= -R(x-y_{r_0}) + \frac{Q y}{2\pi (x^2 + y^2)}, \label{eq:vortex_sink_v}
\end{align}
\end{subequations}
where $R = 0.9$ is a scalar parameter controlling the strength and direction of the vortex, $Q = -1.57$ is the strength of the local sink, and $(x_{r_0},\, y_{r_0}) = (0,\, 0)$ is the centre of rotation. This flow diverges as $x,y\to\pm\infty$, so we include an additional multiplicative term to decay the solvent velocity to zero such that $\tilde{\vec{u}}(x,y) = \vec{u}(x,y)e^{-(x^2+y^2)/\alpha}$, with the choice $\alpha = 8.5$. We have maintained the same external potential width $\sigma$, time scale $\tau$, friction coefficient $\gamma$, as well as the number of particles $N = 50$ from the previous example, and once again we have used \citet{rosenfeld1989free} FMT as the excess over ideal gas free energy functional $\mathcal{F}_{ex}$. For this example we use a slowly decaying confining potential 
\begin{align}
V_1(x,y,t)= \lambda(x^2+y^2) -\lambda e^{-\left(\frac{t}{\tau}\right)^2}
e^{-\left(\frac{(x-x_a)^2}{\sigma^2} + \frac{(y-y_a)^2}{\sigma^2}\right)},
\label{eq:V1_fig2}
\end{align}
with  $\lambda = 0.001$ and $(x_a,\,y_a) = (0,\,4) $. Here, $V_1(x,y,t) \to \lambda(x^2+y^2)$ as $t\to \infty$, which is a weak background potential necessary to maintain a nonzero density over the infinite 2D plane. In Figure \ref{fig:sink_flow} we plot time snap-shots of the numerical solution. A non-trivial density evolution due to the combined effect of inertia and the external flow may be observed. We observe in the $\vec{u} \neq 0$ case on the right that the flow field \eqref{eq:vortex_sink_u}--\eqref{eq:vortex_sink_v} drags the colloids around in a swirling circular motion. Eventually, in the steady state, the effects of hard sphere exclusion become apparent, with the colloids forming a stable ring shaped density distribution around the sink location. 

\subsection{Flow in a Periodic Strip}

To illustrate the effects of externally flowing solvents on the dynamics of the density combined with geometrical confinement we consider a system of hard spheres, of radius 0.5, confined in a channel of width 4 ($x \in [-2,2]$) and length $2\pi$ (between $y \in [0,2\pi]$).  The domain is periodic in the $y$-direction, and we impose no-flux boundary conditions on the density and current in the $x$-direction. In other words, $\vec{j}\cdot \vec{e}_x = 0$ on $x = -2$, $x = 2$ and $t\geq 0$, and $\varrho(\vec{r},t) = \varrho(\vec{r} + 2\pi\vec{e}_y,t)$ for $(x,y)\in[-2,2]\times[0,2\pi]$ and $t \geq 0$. We impose a parabolic external flow of the form
\begin{subequations}
\begin{align}
	u &= 0, \\
	v &= A(t)(y-W/2)(y+W/2),
\end{align}
\end{subequations}
where $W=4$ is the width of the channel, and $A(t)$ controls the strength and direction of the flow via:
\[
	A(t) = \begin{cases} 
	0, \quad t = 0 \\
	\alpha, \quad t \in (0,t_s] \\
	-\alpha, \quad t > t_s
	\end{cases}
\]
where $\alpha = 1$ and $t_s = 0.5$.  This represents a parabolic flow which (instantaneously) switches direction at a given time ($t_s$).

We choose an initial condition of the form
\[
	\rho = Z( \cos( y + y_0 ) + \rho_0),
\]
where $y_0 = \pi$ and $\rho_0 = 2$.  We choose $Z$ such that $\int \rho d x d y = 4$.  Note that this is not the total number of particles in the system, but rather a measure of the (average) area of the 2D slice containing particles; the true number of particles in the 3D system can be arbitrarily large.

In Figure~\ref{fig:channelFlow} we clearly see significant effects of both the channel walls (which causes non-trivial packing of particles at the walls; note that we do not show the excluded-volume region where the density is zero), and inertia. The latter is demonstrated both at times before the switch of the external flow (at $t=0.4$, it can be seen that the peak of the underdamped distribution lags behind that of the overdamped dynamics), and shortly after the switch of the external flow (at $t=0.6$), where the velocity predicted by the DDFT incorporating inertial effects remains in the opposite direction to the external flow. In contrast, the flux of the overdamped dynamics instantaneously switches direction at $t=0.5$. For longer times, the velocities of the inertial system align with the external flow, until, for very long times, the two densities become indistinguishable, where the underdamped velocity aligns with the external flow (not shown).

\section{Summary and Outlook}\label{sec:discussion}

We have derived a general DDFT for systems of colloidal particles, including the effects of inertia, general time-dependent, nonhomogeneous externally flowing solvents, and hydrodynamic interactions. Additionally, in the overdamped limit, the total free energy in the system may be shown to be decreasing along the external flow. For the trivial flow field $\vec{u} = 0$, the DDFT relaxes towards the equilibrium density given by the free energy functional used for the dynamics. The \emph{ab initio} derivation of the presented DDFT, relies on three assumptions for the novel inclusion of general nonhomogeneous external flows: (i) the externally flowing solvent is incompressible; (ii) in the case of including HI, the $n$--body distributions governing the HI terms remain well approximated by functionals of $\varrho(\vec{r},t)$ and $\vec{v}(\vec{r},t)$; (iii) the density of the particle systems is relatively low. For the latter, in the overdamped limit, \citet{almenar2011dynamics} have shown that the steady state sum rule \eqref{eq:excess_free_energy_equilibrium_sum} is violated in dense systems, by computing the left hand side using the steady state solution to their DDFT and comparing with the right hand side computed with the equilibrium probability density $P (\vec{r}_1 , \vec{r}_2)$ obtained from the the Fokker--Planck equation for the two-point correlation function for a system of exactly two particles. It is likely that the local steady state approximation of the inertial DDFT presented here also breaks down for dense non--equilibrium steady states. To get around this, one could consider closing the hierarchy and obtaining a steady state relation for the three--point correlation function \cite{tschopp2022first, tschopp2023superadiabatic}, or by using the methods of power functional theory \cite{schmidt2022power}. 

For the remainder of the derivation, we rely on the usual approximations: (i) The non-steady contributions from the one-body distribution which are not captured by the local-steady approximation are small or may be neglected.
(ii) For two-body HI, it is assumed that the widely used Enskog approximation is compatible with externally flowing solvents which are subject to the aforementioned assumptions. Further, by neglecting HI and external flows we recover the DDFT derived in Ref.~\onlinecite{archer2009dynamical}, and by neglecting external flows only we recover the DDFT derived in Ref.~\onlinecite{goddard2012unification}. 

Via the numerical solutions presented in Section \ref{sec:numerical_solutions}, we have demonstrated that the inclusion of inertia as well as general, inhomogeneous external flow fields may be combined with the usual approximations made in deriving a DDFT to produce accurate models for colloidal fluids undergoing external forcing, applicable to a wide range of non-trivial systems. Promising future extensions to the presented DDFT, both theoretical and numerical, include the extension to multiple--particle species, anisotropic particles, self-propelled particles, as well as a more thorough investigation of the effects of confined geometries. 

\section*{Data Availability Statement}
The data that support the findings of this study are available from the corresponding author upon reasonable request.

\section*{Author Declarations}
The authors have no conflicts to disclose.

\begin{acknowledgments}
B. D. Goddard would like to acknowledge support from EPSRC EP/L025159/1. R. D. Mills-Williams is grateful to EPSRC for PhD funding. Both B.~D.~Goddard and A.~J.~Archer gratefully acknowledge support for parts of this work from the London Mathematical Society and the Loughborough University Institute of Advanced Studies. 
\end{acknowledgments}

\appendix

\bibliography{aipsamp}

\end{document}